\input harvmac

\Title{\vbox{\baselineskip12pt
\hbox{QMW-PH-99-17}
\hbox{BCUNY-HEP/99-02}
\hbox{hep-th/9911082}}}
{\vbox{\centerline{Worldvolume Theories, Holography, Duality  and Time}}}

\baselineskip=12pt
\centerline {Chris M. Hull$^1$\footnote{$^a$}{C.M.Hull@qmw.ac.uk}
 and Ramzi R. Khuri$^{2,3}$\footnote{$^b$}{khuri@gursey.baruch.cuny.edu}}
\medskip
\centerline{\sl $^1$Department of Physics}
\centerline{\sl Queen Mary and Westfield College}
\centerline{\sl Mile End Road}
\centerline{\sl London E1 4NS UK}
\medskip
\centerline{\sl $^2$Department of Natural Sciences}
\centerline{\sl Baruch College, CUNY}
\centerline{\sl 17 Lexington Avenue}
\centerline{\sl New York, NY 10010}
\centerline{and}
\centerline{\sl The Graduate School and University Center, CUNY}
\centerline{\sl 33 West 42nd Street}
\centerline{\sl New York, NY 10036-8099}
\medskip
\centerline{\sl $^3$\footnote{$^c$}{Associate member}
Center for Advanced Mathematical Sciences}
\centerline{\sl American University of Beirut}
\centerline{\sl Beirut, Lebanon}
\vfill
\eject

\bigskip
\centerline{\bf Abstract}
\medskip
\baselineskip = 20pt

Duality transformations involving compactifications on timelike as well as
spacelike circles link
M-theory,  the 10+1-dimensional strong coupling limit of IIA string theory, to
other
11-dimensional theories in signatures 9+2 and 6+5 and to type II string
theories
in all 10-dimensional signatures.
These theories have  BPS branes of various world-volume signatures, and here we
construct the
world-volume theories for these branes, which in each case have 16
supersymmetries.
For the generalised D-branes of the various type II string theories, these
are always supersymmetric Yang-Mills theories with
16 supersymmetries, and we show that these all arise from compactifications of
the
supersymmetric Yang-Mills theories in 9+1 or 5+5 dimensions.
We discuss the geometry of the
brane solutions
and, for the cases in which the world-volume theories are superconformally
invariant,
we propose holographically dual string or M theories in constant curvature
backgrounds. For product space solutions $X\times Y$, there is in general a
conformal
field theory associated with the boundary of $X$ and another with the boundary
of $Y$.

\Date{November 1999}

\font\mybb=msbm10 at 10pt
\def\bbbb#1{\hbox{\mybb#1}}

\def\R {\bbbb{R}}

\def\({\left (}
\def\){\right )}
\def\[{\left [}
\def\]{\right ]}

\def \aa {\alpha}
\def \bb {\beta}

\def \rr {\rho}
\def \ss {\sigma}
\def \tt {\tau}

\def\sym {super Yang-Mills}
\def \ti {\tilde}

\def \2 {{1 \over 2}}
\def \3 {{1 \over 3}}
\def \4 {{1 \over 4}}
\def \5 {{1 \over 5}}
\def \6 {{1 \over 6}}
\def \7 {{1 \over 7}}
\def \8 {{1 \over 8}}
\def \9 {{1 \over 9}}
\def \0 { \infty}

\def\++ {{(+)}}
\def \- {{(-)}}
\def\+-{{(\pm)}}

\def\ek {\eqn\abc}

\def \pa {\partial}

\def \qq {\qquad}

\def\unit{\hbox to 3.3pt{\hskip1.3pt \vrule height 7pt width .4pt
\hskip.7pt
\vrule height 7.85pt width .4pt \kern-2.4pt
\hrulefill \kern-3pt
\raise 4pt\hbox{\char'40}}}

\def\nup#1({Nucl.\ Phys.\  {\bf B#1}\ (}

\def\s{\sigma}

\lref\chamblin{A. Chamblin and R. Emparan, Phys. Rev. {\bf D55}
(1997) 754.}

\lref\emparan{R. Emparan, Nucl. Phys. {\bf B490} (1997) 365.}

\lref\guven{R. Guven, Phys. Lett. {\bf B276} (1992) 49.}

\lref\bko{C.M. Hull, Phys. Lett. {\bf B39} (1984) 139;
E. A. Bergshoeff, R. Kallosh and T. Ortin,
Phys. Rev. {\bf D47} (1993) 5444.}

\lref\bgppr{J. Barrett, G. W. Gibbons, M. J. Perry, C. N. Pope
and P. Ruback, Int. J. Mod. Phys. {\bf A9} (1994) 1457.}

\lref\eleven{M. J. Duff, P. S. Howe, T. Inami and K. Stelle, Phys. Lett.
{\bf B191} (1987) 70.}

\lref\stain{H. Lu, C. N. Pope, E. Sezgin, and K. S. Stelle,
Nucl. Phys. {\bf B456} (1995) 669; see also H. Lu, C. N. Pope,
and K. S. Stelle, Nucl. Phys. {\bf B481} (1996) 313 and references
therein.}

\lref\ddduff{M. J. Duff and J. X. Lu, Nucl. Phys. {\bf B416} (1994) 301.}

\lref\chrisone{C. M. Hull, {\bf JHEP} 9807:021, 1998, hep-th/9806146.}

\lref\christwo{C. M. Hull, {\bf JHEP} 9811:017, 1998, hep-th/9807127.}

\lref\Crev{C. M. Hull,  hep-th/9911080.}

\lref\us{C. M. Hull and R. R. Khuri, Nucl. Phys. {\bf B536} (1999) 219,
hep-th/9808069.}

\lref\mal{J. Maldacena, hep-th/9711200.}

\lref\nextpaper{C. M. Hull and R. R. Khuri, in preparation.}


\lref\rust{R. R. Khuri and R. C. Myers, Nucl. Phys. {\bf B466}
(1996) 60.}

\lref\GSW {M. B. Green, J. H. Schwarz and E. Witten,
{\it Superstring Theory}, Cambridge University Press, Cambridge (1987).}

\lref\prep{See M. J. Duff, R. R. Khuri and J. X. Lu, Phys. Rep.
{\bf B259} (1995) 213 and references therein.}

\lref\PKT{P.K. Townsend, Phys. Lett. {\bf B350}  (1995) 184.}

\lref\moore{G. Moore, hep-th/9305139,9308052.}

\lref\CJ{E. Cremmer and B. Julia, Phys. Lett. {\bf 80B} (1978) 48; Nucl.
Phys. {\bf B159} (1979) 141.}

\lref\julia{B. Julia in {\it Supergravity and Superspace}, S.W. Hawking
and M. Ro$\check c$ek, C.U.P.
Cambridge,  (1981). }

\lref\julec{B.  Julia, hep-th/9805083.}

\lref\HT{C.M. Hull and P.K. Townsend, Nucl. Phys. {\bf B438} (1995) 109,
hep-th/9410167.}

\lref\gibrap{G.W. Gibbons, hep-th/9803206.}

\lref\buscher{T. H. Buscher, Phys. Lett. {\bf 159B} (1985) 127,
Phys. Lett. {\bf B194}
 (1987), 51 ; Phys. Lett. {\bf B201}
 (1988), 466.}

\lref\bergort{E. Bergshoeff, C.M. Hull and T. Ortin,
 Nucl. Phys. {\bf B451} (1995) 547,hep-th/9504081.}

\lref\rocver {M. Ro\v cek and E. Verlinde, Nucl. Phys.
{\bf B373} (1992), 630.}

 \lref\givroc {A. Giveon, M.
Ro\v cek, Nucl. Phys. {\bf B380} (1992), 128.}

\lref\alv{E. Alvarez, L. Alvarez-Gaum\' e,
J.L. Barbon and Y. Lozano,  Nucl. Phys. {\bf B415} (1994)
71.}

\lref\TD {A. Giveon, M. Porrati and E. Rabinovici, Phys. Rep. {\bf 244}
(1994) 77.}

\lref\HJ{C. M. Hull  and B. Julia, hep-th/9803239.}

\lref\CPS{E. Cremmer,  I.V. Lavrinenko,  H. Lu,
C.N. Pope,  K.S. Stelle and  T.A. Tran, hep-th/9803259.}

\lref\Stelle{
K. S. Stelle, hep-th/9803116.}

\lref\ddua {J. Dai, R.G. Leigh and J. Polchinski, Mod. Phys. Lett. {\bf
A4} (1989) 2073.}

\lref\dsei{ M. Dine, P. Huet and N. Seiberg, Nucl. Phys. {\bf B322}
(1989) 301.}

\lref\gibras{  G.W. Gibbons and D.A. Rasheed, hep-th/904177.}

\lref\Bob{B.S. Acharya, M. O'Loughlin and B. Spence, Nucl.
Phys. {\bf B503} (1997) 657; B.S. Acharya, J.M. Figueroa-O'Farrill, M.
O'Loughlin and B. Spence,
hep-th/9707118.}

\lref\thom{M. Blau and G. Thompson, Phys. Lett. {\bf B415} (1997) 242.}

\lref\hawtim{S.W. Hawking,  Phys.Rev. {\bf D46 } (1992) 603.}

\lref\dinst{M.B.  Green, Phys. Lett. {\bf  B354}
(1995) 271,  hep-th/9504108;
M.B.~Green and M.~Gutperle,   Phys.Lett. B398(1997)69, hep-th/9612127;
M.B.~Green and M.~Gutperle,   hep-th/9701093,
G.~Moore, N.~Nekrasov and  S.~Shatahvilli,   hep-th/9803265;
E. Bergshoeff and  K. Behrndt, hep-th/9803090.}

\lref\sevbrane{G.W. Gibbons, M.B.  Green and M.J. Perry, Phys.Lett.
B370 (1996) 37, hep-th/9511080.}

\lref\beck{K.~Becker, M.~Becker and  A.~Strominger, hep-th/9507158,
Nucl.Phys. {\bf 456} (1995) 130.}

\lref\dinstcal{K.~Becker, M.~Becker, D.R.~Morrison, H.~Ooguri, Y.~Oz and
Z.~Yin,
  hep-th/9608116, Nucl.Phys. {\bf 480} (1996) 225; H.~Ooguri and C.~Vafa,
hep-th/9608079, Phys. Rev. Letts. 77(1996) 3296; M.~Gutperle,
hep-th/9712156.}

\lref\adsstab{P. Breitenlohner and D.Z. Freedman, Phys. Lett. {\bf 115B}
(1982) 197; Ann. Phys. {\bf 144} (1982) 197; G.W. Gibbons, C.M. Hull and
N.P. Warner, Nucl. Phys. {\bf B218} (1983) 173.}

\lref\OS{K. Osterwalder and R. Schrader, Phys. Rev. Lett.
{\bf 29} (1972) 1423; Helv. Phys. Acta{\bf 46}
 (1973) 277;
CMP {\bf 31} (1973) 83 and CMP {\bf 42} (1975) 281.
K. Osterwalder,  in G. Velo and A. Wightman (Eds.)
Constructive Field Theory - Erice lectures 1973,
Springer-Verlag Berlin 1973; K. Osterwalder in {\it Advances in Dynamical
Systems and
Quantum Physics}, Capri conference, World Scientific 1993. }

\lref\vanwick{P. van Nieuwenhuizen and A. Waldron, Phys.Lett. B389 (1996)
29-36, hep-th/9608174. }

\lref\KT{T. Kugo and P.K. Townsend, Nucl. Phys. {\bf B221} (1983) 357.}

\lref\yam{J. Yamron, Phys. Lett. {\bf B213} (1988) 325.}

\lref\vafwit{C. Vafa and E. Witten, Nucl. Phys. {\bf B431} (1994) 3-77,
hep-th/9408074.}

\lref\sing{ L. Baulieu, I. Kanno and I. Singer, hep-th/9704167.}

\lref\Seib{N. Seiberg,
hep-th/9705117.}%

\lref\vans{K. Pilch, P. van Nieuwenhuizen and M. Sohnius, Commun.
Math. Phys.
{\bf 98} (1985) 105.}

\lref\luk{J. Lukierski and A. Nowicki,   Phys.Lett. {\bf 151B}
(1985)  382.}

\lref\witt{E. Witten, Adv. Theor. Math. Phys. {\bf 2} (1998)
253, hep-th/9802150.}

\lref\holog{L. Susskind and E. Witten, hep-th/9805114;
S.S. Gubser, I.R. Klebanov and A.M. Polyakov, Phys. Lett. {\bf B428}
(1998) 105, 9801003.}

\lref\bergort{E. Bergshoeff, C.M. Hull and T. Ortin, Nucl. Phys.
{\bf B451}
(1995) 547, hep-th/9504081.}

\lref\huto{C.M. Hull, Nucl. Phys. {\bf B509} (1998) 252,
hep-th/9702067. }

\lref\asp{P. Aspinwall, Nucl. Phys. Proc. Suppl. {\bf  46}  (1996) 30,
hep-th/9508154; J. H. Schwarz, hep-th/9508143.}

\lref\fvaf{C. Vafa, Nucl. Phys. {\bf 469} (1996) 403.}

\lref\GravDu{C.M. Hull, Nucl. Phys. {\bf B509} (1997) 252,
hep-th/9705162.}
\lref\bergnin{ E.~Bergshoeff and J.P.~van der Schaar,
                {  hep-th/9806069}.}

\lref\ythe{C.M. Hull, Nucl.Phys. {\bf B468}  (1996) 113  hep-th/9512181;
A. A. Tseytlin Nucl.Phys.
{\bf B469}  (1996) 51  hep-th/9602064.}

\lref\mythe{I. Bars,  Phys. Rev. {\bf D54} (1996) 5203, hep-th/9604139;
hep-th/9604200;
Phys.Rev. {\bf D55} (1997) 2373 hep-th/9607112 .}

\lref\huku{C.M. Hull and R.R. Khuri, in preparation.}

\lref\inter{G. W. Gibbons and P. K. Townsend, Phys. Rev. Lett.
{\bf 71} (1993) 3754; M. J. Duff, G. W. Gibbons and P. K. Townsend,
Phys. Lett. {\bf B332} (1994) 321; G. W. Gibbons, G. T. Horowitz and
P. K. Townsend, Class. Quan. Grav. {\bf 12} (1995) 297.}

\lref\adssol{M. Gunaydin and N. Marcus, Class. Quant. Grav {\bf 2}
(1985) L11;
 H.J. Kim, L.J. Romans
and P. van Nieuwenhuizen, Phys. Rev. {\bf D32} (1985) 389.}

\lref\CW{C.M. Hull and N. P. Warner, Class. Quant. Grav. {\bf 5}
(1988) 1517.}

\lref\poly{S.S. Gubser,  I. R. Klebanov and  A. M. Polyakov,
hep-th/9802109.}

\lref\hor{ G. Horowitz and A. Strominger, Nucl. Phys {\bf B360}
(1991) 197.}

\lref\huten{C.M. Hull, Phys. Lett. {\bf B357 } (1995) 545,
hep-th/9506194.}

\lref\gibhaw{G.W. Gibbons and S.W. Hawking, Phys. Rev. {\bf D15}
(1977) 2738.}

\lref\kall{P. Claus, R. Kallosh, J. Kumar, P.K. Townsend and A. van
Proeyen,
hep-th/9801206.}

\lref\witkk{E. Witten, Nucl. Phys. {\bf B195} (1982) 481.}

\lref\haweuc{S.W. Hawking, in {\it General Relativity}, ed. by S.W.
Hawking
and W. Israel, Cambridge University Press, 1979.}

\lref\dufstel{M.J. Duff and K.S. Stelle, Phys. Lett. {\bf B253}
(1991) 113.}

\lref\horwit{P. Horava and E. Witten, Nucl. Phys. {\bf B460} (1996) 506.}

\lref\wit{E. Witten, Nucl. Phys. {\bf B443} (1995) 85.}

\lref\blencowe{M. P. Blencowe and M. J. Duff, Nucl. Phys.
{\bf B10} (1988) 387.}

\lref\fr{P. G. O. Freund and M. A. Rubin, Phys. Lett. {\bf B97}
(1980) 233;
F. Englert,  Phys. Lett. {\bf B119} (1982) 339; see also M. J. Duff,
B. E. W. Nilsson and C. N. Pope, Phys. Rep. {\bf 130} vols. 1 \& 2
(1986) 1 and references therein.}

\lref\witads{E. Witten, hep-th/9802150.}

\lref\pvt{K. Pilch, P. van Nieuwenhuizen and P. K. Townsend,
Nucl. Phys. {\bf B242} (1984) 377.}

\lref\cjs{E. Cremmer, B. Julia and J. Scherk,
Phys. Lett.{\bf B76} (1978) 409.}

\lref\gps{D. J. Gross and M. J. Perry, Nucl. Phys. {\bf B226} (1983) 29;
Phys. Rev. Lett. {\bf 51} (1983) 87.}

\lref\seibergir{N. Seiberg, Phys. Lett. {\bf B384} (1996) 81.}


%
%
%
%
\newhelp\stablestylehelp{You must choose a style between 0 and 3.}%
\newhelp\stablelinehelp{You should not use special hrules when stretching%
a table.}%
\newhelp\stablesmultiplehelp{You have tried to place an S-Table %
inside another%
S-Table.  I would recommend not going on.}%
%
%
\newdimen\stablesthinline
\stablesthinline=0.4pt
\newdimen\stablesthickline
\stablesthickline=1pt
%
%
\newif\ifstablesborderthin
\stablesborderthinfalse
\newif\ifstablesinternalthin
\stablesinternalthintrue
\newif\ifstablesomit
\newif\ifstablemode
\newif\ifstablesright
\stablesrightfalse
%
%
\newdimen\stablesbaselineskip
\newdimen\stableslineskip
\newdimen\stableslineskiplimit
%
%
\newcount\stablesmode
\newcount\stableslines
\newcount\stablestemp
\stablestemp=3
\newcount\stablescount
\stablescount=0
\newcount\stableslinet
\stableslinet=0
%
%
%
\newcount\stablestyle
\stablestyle=0
%
%
\def\stablesleft{\quad\hfil}%
\def\stablesright{\hfil\quad}%
%
%
\catcode`\|=\active%
%
%
\newcount\stablestrutsize
\newbox\stablestrutbox
\setbox\stablestrutbox=\hbox{\vrule height10pt depth5pt width0pt}
\def\stablestrut{\relax\ifmmode%
                         \copy\stablestrutbox%
                       \else%
                         \unhcopy\stablestrutbox%
                       \fi}%
%
%
\newdimen\stablesborderwidth
\newdimen\stablesinternalwidth
\newdimen\stablesdummy
\newcount\stablesdummyc
\newif\ifstablesin
\stablesinfalse
%
%
\def\begintable{\stablestart%
  \stablemodetrue%
  \stablesadj%
  \halign%
  \stablesdef}%
\def\stablesadj{%
  \ifcase\stablestyle%
    \hbox to \hsize\bgroup\hss\vbox\bgroup%
  \or%
    \hbox to \hsize\bgroup\vbox\bgroup%
  \or%
    \hbox to \hsize\bgroup\hss\vbox\bgroup%
  \or%
    \hbox\bgroup\vbox\bgroup%
  \else%
    \errhelp=\stablestylehelp%
    \errmessage{Invalid style selected, using default}%
    \hbox to \hsize\bgroup\hss\vbox\bgroup%
  \fi}%
\def\stablesend{\egroup%
  \ifcase\stablestyle%
    \hss\egroup%
  \or%
    \hss\egroup%
  \or%
    \egroup%
  \or%
    \egroup%
  \else%
    \hss\egroup%
  \fi}%
\def\stablestart{%
  \ifstablesin%
    \errhelp=\stablesmultiplehelp%
    \errmessage{An S-Table cannot be placed within an S-Table!}%
  \fi
  \global\stablesintrue%
  \global\advance\stablescount by 1%
  \message{<S-Tables Generating Table \number\stablescount}%
  \begingroup%
  \stablestrutsize=\ht\stablestrutbox%
  \advance\stablestrutsize by \dp\stablestrutbox%
  \ifstablesborderthin%
    \stablesborderwidth=\stablesthinline%
  \else%
    \stablesborderwidth=\stablesthickline%
  \fi%
  \ifstablesinternalthin%
    \stablesinternalwidth=\stablesthinline%
  \else%
    \stablesinternalwidth=\stablesthickline%
  \fi%
  \tabskip=0pt%
  \stablesbaselineskip=\baselineskip%
  \stableslineskip=\lineskip%
  \stableslineskiplimit=\lineskiplimit%
  \offinterlineskip%
  \def\borderrule{\vrule width \stablesborderwidth}%
  \def\internalrule{\vrule width \stablesinternalwidth}%
  \def\thinline{\noalign{\hrule height \stablesthinline}}%
  \def\thickline{\noalign{\hrule height \stablesthickline}}%
  \def\trule{\omit\leaders\hrule height \stablesthinline\hfill}%
  \def\ttrule{\omit\leaders\hrule height \stablesthickline\hfill}%
  \def\tttrule##1{\omit\leaders\hrule height ##1\hfill}%
  \def\stablesel{&\omit\global\stablesmode=0%
    \global\advance\stableslines by 1\borderrule\hfil\cr}%
  \def\el{\stablesel&}%
  \def\elt{\stablesel\thinline&}%
  \def\eltt{\stablesel\thickline&}%
  \def\elttt##1{\stablesel\noalign{\hrule height ##1}&}%
  \def\elspec{&\omit\hfil\borderrule\cr\omit\borderrule&%
              \ifstablemode%
              \else%
                \errhelp=\stablelinehelp%
                \errmessage{Special ruling will not display properly}%
              \fi}%
  \def\stmultispan##1{\mscount=##1 \loop\ifnum\mscount>3 \stspan\repeat}%
  \def\stspan{\span\omit \advance\mscount by -1}%
  \def\multicolumn##1{\omit\multiply\stablestemp by ##1%
     \stmultispan{\stablestemp}%
     \advance\stablesmode by ##1%
     \advance\stablesmode by -1%
     \stablestemp=3}%
  \def\multirow##1{\stablesdummyc=##1\parindent=0pt\setbox0\hbox\bgroup%
    \aftergroup\emultirow\let\temp=}
  \def\emultirow{\setbox1\vbox to\stablesdummyc\stablestrutsize%
    {\hsize\wd0\vfil\box0\vfil}%
    \ht1=\ht\stablestrutbox%
    \dp1=\dp\stablestrutbox%
    \box1}%
  \def\stpar##1{\vtop\bgroup\hsize ##1%
     \baselineskip=\stablesbaselineskip%
     \lineskip=\stableslineskip%

\lineskiplimit=\stableslineskiplimit\bgroup\aftergroup\estpar\let\temp=}%
  \def\estpar{\vskip 6pt\egroup}%
  \def\stparrow##1##2{\stablesdummy=##2%
     \setbox0=\vtop to ##1\stablestrutsize\bgroup%
     \hsize\stablesdummy%
     \baselineskip=\stablesbaselineskip%
     \lineskip=\stableslineskip%
     \lineskiplimit=\stableslineskiplimit%
     \bgroup\vfil\aftergroup\estparrow%
     \let\temp=}%
  \def\estparrow{\vfil\egroup%
     \ht0=\ht\stablestrutbox%
     \dp0=\dp\stablestrutbox%
     \wd0=\stablesdummy%
     \box0}%
  \def|{\global\advance\stablesmode by 1&&&}%
  \def\|{\global\advance\stablesmode by 1&\omit\vrule width 0pt%
         \hfil&&}%
  \def\vt{\global\advance\stablesmode by 1&\omit\vrule width
\stablesthinline%
          \hfil&&}%
  \def\vtt{\global\advance\stablesmode by 1&\omit\vrule width
\stablesthickline%
          \hfil&&}%
  \def\vttt##1{\global\advance\stablesmode by 1&\omit\vrule width ##1%
          \hfil&&}%
  \def\vtr{\global\advance\stablesmode by 1&\omit\hfil\vrule width%
           \stablesthinline&&}%
  \def\vttr{\global\advance\stablesmode by 1&\omit\hfil\vrule width%
            \stablesthickline&&}%
\def\vtttr##1{\global\advance\stablesmode by 1&\omit\hfil\vrule width ##1&&}%
\stableslines=0%
\stablesomitfalse}%
\def\stablesdef{\bgroup\stablestrut\borderrule##\tabskip=0pt plus 1fil%
  &\stablesleft##\stablesright%
  &##\ifstablesright\hfill\fi\internalrule\ifstablesright\else\hfill\fi%
  \tabskip 0pt&&##\hfil\tabskip=0pt plus 1fil%
  &\stablesleft##\stablesright%
  &##\ifstablesright\hfill\fi\internalrule\ifstablesright\else\hfill\fi%
  \tabskip=0pt\cr%
  \ifstablesborderthin%
    \thinline%
  \else%
    \thickline%
  \fi&%
}%
\def\endtable{\advance\stableslines by 1\advance\stablesmode by 1%
   \message{- Rows: \number\stableslines, Columns:
\number\stablesmode>}%
   \stablesel%
   \ifstablesborderthin%
     \thinline%
   \else%
     \thickline%
   \fi%
   \egroup\stablesend%
\endgroup%
\global\stablesinfalse}
%
%



\newsec{Introduction}

Duality maps relate the
five distinct perturbative string theories in 9+1 dimensions \refs{\HT,\wit},
and these are now
understood
as different   limits of a   theory
in 10+1 dimensions, the so-called M-theory. For the purposes of this paper, we
will define M-theory
as the 10+1 dimensional theory arising as the strong-coupling limit of the IIA
string theory.
In \refs{\chrisone,\christwo}, this picture was expanded
to include  dualities involving compactification on timelike circles as well as
spacelike ones. In
\chrisone, it was shown that T-duality on a time-like circle takes the IIA
theory into a IIB* theory
and the IIB theory into a IIA* theory.
The strong coupling limit of the IIA* theory is a theory in 9+2 dimensions,
denoted the M* theory
in \christwo, which can also be obtained by compactifying M-theory on a
Lorentzian 3-torus
$T^{2,1}$ with 2 spacelike circles and one timelike one, and taking the limit
in which all three
circles shrink to zero size.
Compactifying the M* theory on a Euclidean 3-torus $T^3$ and shrinking
the
torus to zero size then
gives an M$'$ theory in 6+5 dimensions. Compactifying the M, M*, M$'$
theories on spacelike or timelike
circles gives rise to IIA-like string theories in signatures 10+0,
9+1, 8+2, 6+4 and 5+5, and
T-dualities relate these to IIB-like string theories in signatures 9+1,
7+3 and 5+5.
Each of these theories has 32 local supersymmetries, which in some cases
satisfy a twisted
supersymmetry algebra \christwo, and each has a supergravity limit.  These
theories are linked by an intricate web of duality
transformations which can change the number of time dimensions as well as the
number of space
dimensions.
As all of these theories are linked by dualities, they should all be regarded
as different limits
of a single underlying theory. For a review, see \Crev.

In \us, the generalised brane-type solutions of these theories
were constructed,
and found to
have various world-volume signatures. For example, the 9+2 dimensional
M* theory has membrane-type solutions with world-volumes of signature (3,0)
and (1,2) and a solitonic fivebrane-type solution with signature (5,1).
The rules for determining which signature branes appear in which theories
were given in \us, as well as the
duality transformations relating
 the various
generalised  branes. These were found to be
consistent with considerations of dimensional reduction from 11 to 10
dimensions.

There are of course many issues arising concerning the interpretation of these
theories with
multiple times, some of which are discussed in
\refs{\chrisone,\christwo,\Crev}. Here
we will proceed
formally and continue to map out the structure of the theories that arise.

The plan of this paper is as follows: in section 2, we review the D-brane and
E-brane solutions of
the type II and type II* string theories and discuss their singularities. In
section 3, we review
some of the main results of \us,
emphasizing the various brane solutions obtained in  10 and 11 dimensions.
In section 4, we interpret the branes as interpolating solutions,
summarizing the results in eleven dimensions found in \us\ and explicitly
writing the interpolating geometries for the family of four-dimensional brane
solutions of the
ten-dimensional type IIB theories. In section
5, we discuss world-volume actions for these generalized branes, noting
that each can be linked to a standard brane of M-theory. In
sections 6,7 and 8 we discuss the world-volume theories of the D-branes
of the various type II theories, and the M-branes of the M-type theories.
Finally, in section 9 we discuss the
dualities between conformal field theories and de Sitter space theories
following the arguments of Maldacena \mal.

Various
 generalised de Sitter spaces of
various signatures arise in solutions of these theories \us,
all of which are coset spaces $SO(p,q)/SO(p-1,q)$ with the $SO(p,q)$-invariant
metric and signature $(p-1,q)$; when these
have two sheets, we take one connected component.
 These include $d$-dimensional
de Sitter space
\eqn\eka{dS_d={SO(d,1) \over SO(d-1,1)}{ } ,}
$d$-dimensional anti-de Sitter space
\eqn\ekb{AdS_d={SO(d-1,2) \over SO(d-1,1)}{ },}
the
$d$-sphere
\eqn\ekc{S^d={SO(d+1) \over SO(d)}{ },}
the $d$-hyperboloid
\eqn\ekd{H^d={SO(d,1) \over SO(d)}}
(which has a Euclidean metric and was referred to in \witads\ as
Euclidean anti-de Sitter space) and the space
\eqn\eke{AAdS_d={SO(d-1,2) \over SO(d-2,2)}{},}
with two-timing signature $(d-2,2)$.

\newsec{D-Branes and E-Branes}

The  IIA* and IIB* theories in 9+1 dimensions contain E-branes \chrisone,
which are the
images of D-branes under timelike T-duality.
Whereas D-branes are timelike planes on which strings can end, the
E-branes are spacelike surfaces on which strings can end.
The strings ending on the E-branes govern the behaviour of the E-branes, and
the zero-slope limit
of this string theory gives world-volume theories which are
 Euclidean
super-Yang-Mills theories obtained by reducing 9+1 dimensional
super-Yang-Mills on Lorentzian tori. In particular, the E4-brane solution of
the IIB* theory leads, following \mal, to a duality between the large
N limit of {\it Euclidean} 4-dimensional $U(N)$ super-Yang-Mills theory and
the IIB* string on the product of the 5-dimensional de Sitter space $dS_5$ and
the
hyperbolic 5-space $H_5$  \chrisone.

The bosonic part of the IIA supergravity action is
\eqn\twoa{S_{IIA}=\int d^{10} x \sqrt{-g}\left[
 e^{-2 \Phi}\left(R+ 4(\partial   \Phi )^2  -{H^2\over 12}\right)
-{G_2^2\over 4} - {G_4^2\over 48} \right] +
{4\over \sqrt 3}\int G_4 \wedge G_4 \wedge B_2 + \dots }
while that of IIB supergravity is
\eqn\twob{S_{IIB}=\int d^{10} x \sqrt{-g}\left[
 e^{-2 \Phi} \left( R+ 4(\partial   \Phi )^2
- {H^2\over 12} \right)-{G_1^2\over 2} -  {G_3^2\over 12}
- {G_5^2\over 240}\right] + \dots}
Here  $\Phi $ is the dilaton, $H=dB_2$ is the field strength of the
NS-NS 2-form gauge field $B_2$ and
$G_{n+1}=dC_n+\dots$ is the field strength for the RR $n$-form gauge field
$C_n$. The field
equations derived from the IIB action \twob\ are
supplemented with the self-duality constraint $G_5=*G_5$.
Our conventions are that in signature $S+T$, the metric has $S$ positive
spacelike eigenvalues
and $T$ negative timelike ones, so that
a Lorentzian metric has signature $(S,1)$, i.e. $(++\dots +-)$.

The  type II supergravity solution for a D$p$-brane ($p$ is even
for IIA and odd for IIB) is given by
\refs{\hor,\prep,\huten}
\eqn\dbr{\eqalign{
 ds^2&=H^{-1/2}(-dt^2+dx_1^2+\dots+dx_p^2)+H^{1/2}(dy_{p+1}^2+\dots+dy_9^2)\cr
e^{-2\Phi}&=H^{(p-3)/2}, \qquad\qquad C_{012\dots p}=-H^{-1}+k,\cr}}
where $H$ is a harmonic function of the transverse coordinates
$y_{n+1},\dots,y_9$, $k$ is a constant and here and throughout in the
paper we denote longitudinal spatial coordinates by $x_a$ and transverse
spatial coordinates by $y_i$.
The simplest choice for $H$ is
\eqn\hre{H=c+ {Q\over y^{7-p}},}
where $c$ is a constant (which can be taken to be $0$ or $1$), $y$ is
the radial coordinate defined by
\eqn\abc{y^2=\sum_{i=p+1}^9 y_i^2 } and
$Q$ is proportional to the
D-brane tension, and is taken to be positive (taking $Q<0$ gives an
unphysical negative-tension
brane
with a naked singularity where $H=0$).
When $c\neq 0$, it is conventional to set $k=c^{-1}$, so that as
$y \to \infty$, $C_{012\dots p} \to 0$. However, for convenience we will
henceforth set $k=0$ and usually take $c=1$.

The bosonic part of the type IIA* and type IIB* supergravity actions are given
by reversing
the signs of the RR kinetic terms in \twoa,\twob\ to give \chrisone
\eqn\twoas{S_{IIA*}=\int d^{10} x \sqrt{-g}\left[
 e^{-2 \Phi} \left(  R+ 4(\partial   \Phi )^2
-{H^2\over 12} \right)+{G_2^2\over 4} +{G_4^2\over 48}\right] + \dots}
and
\eqn\twobs{S_{IIB*}=\int d^{10} x \sqrt{-g}\left[
 e^{-2 \Phi} \left( R+ 4(\partial   \Phi )^2
- {H^2\over 12}  \right)+{G_1^2\over 2} +  {G_3^2\over 12}
+ {G_5^2\over 240} \right] + \dots,}
where the field equations from \twobs\ are supplemented by the constraint
$G_5=*G_5$. The full theories are invariant under a twisted $N=2$ supersymmetry
\chrisone.

The E$p$-brane solutions to \twoas\ and \twobs\ are given by
\eqn\ebr{\eqalign{
 ds^2&=H^{-1/2}(dx_1^2+\dots+dx_p^2)+H^{1/2}
(-dt^2+dy_{p+1}^2+\dots+dy_9^2),\cr
e^{-2\phi}&=H^{(p-4)/2}, \qquad \qquad C_{12\dots p}=-H^{-1}.\cr}}
In this case, $H$ is a harmonic function of
$t,y_{p+1},\dots,y_9$
(i.e. it is a solution of the
wave equation $\nabla ^2
H=0$), and $H$ can depend on time as
well as the spatial transverse coordinates.

There are a number of different possibilities for $H$. First, we can take the
time-independent $H$ given by \hre.
These are the solutions that arise from the D-brane supergravity solutions on
performing a timelike
T-duality, using a generalisation \christwo\ of  the usual T-duality rules
\buscher,\bergort.
Secondly, we can consider the  solution  \chrisone\
\eqn\ehar{ H=c+ {Q\over \tt ^{8-p}}}
where $\tt, \ss $ are the proper time and distance defined by
\eqn\abcd{\tt^2 =  -\ss ^2=  t^2- y^2.}
This corresponds to a source located at a point in the transverse space-time.
For odd $p$, taking
\eqn\ehars{H=c+ {Q'\over \ss ^{8-p}}}
gives a different real solution, related to \ehar\ by taking $Q'=iQ$.

The  solutions have   potential  singularities on the light-cone $t^2=y^2$,
where $H$ diverges, and on the hyperboloid where $H=0$.
There are in general two distinct solutions, one in which  the
coordinates are restricted to the
interior of the light cone, $t^2 \ge y^2$, and one in which they are
restricted to the
exterior of the light cone, $t^2 \le y^2$. For example, for the E4-brane, it
was shown in
 \chrisone\ that the geometry near $t^2=y^2$ is non-singular and approaches
$dS_5\times H^5$, the
product of 5-dimensional de Sitter space and hyperbolic 5-space with constant
negative-curvature
and positive definite metric. Taking  $Q$ to be positive (with $c=1$) so
that $H$ is non-vanishing
then avoids the potential singularity at $H=0$.
The   region  $t^2 \le y^2$ then
defines a non-singular solution
which interpolates between flat space (the region in which $\sigma$ is large)
and
 $dS_5\times H^5$ (where $\sigma$ is small), and is geodesically complete
\chrisone.
The interior of the light-cone  $t^2 \ge y^2$
also defines a non-singular geodesically complete solution, with the future and
past regions
$t\ge 0$ and $t\le 0$ each giving a coordinate patch covering half the space.
The region near $t^2=y^2$ is again $dS_5\times H^5$, with the future-cone
$t>y>0$
covering one half of $dS^5$ and the past-cone covering the other half
\chrisone.
The situation is similar for the other E-branes, with the interior and
exterior solutions defining two different solutions \refs{\nextpaper}.
For the solution with timelike interpolation in which $t^2\ge y^2$, we take the
harmonic function to be
given by
\ehar\ with $Q>0$, while for  the solution with spacelike interpolation in
which
$t^2\le y^2$, we take the harmonic function to be given by
\ehars\ with $Q'>0$,
  so that in each case
$H>0$.

The solution \ebr\ is an extended object associated with a spacelike
$p$-surface with
coordinates $x^1,\dots ,x^p$
located at $t=y^{p+1}=\dots =y^9=0$.
This is to be compared with  a D-brane, which is associated
 with a timelike $p+1$-surface with coordinates $t,x^1,\dots ,x^p$
located at $ y^{p+1}=\dots =y^9=0$.
A D-brane arises in perturbative type II string theory from imposing
Dirichlet boundary conditions
in the directions $ y^{p+1},\dots ,y^9 $ and Neumann conditions in the
remaining directions, and the
D-brane solution \dbr\ describes the supergravity fields resulting from
such a D-brane source.
Similarly, the E-brane in perturbative type II* string theory
arises from imposing Dirichlet
boundary conditions in the directions $t, y^{p+1},\dots ,y^9 $, including
time, and Neumann conditions
in the remaining directions, and the E-brane solution \ebr\
describes the supergravity fields resulting
from such an E-brane source. In the perturbative string theory, the
timelike T-duality taking the
type II theory to the type II* theory changes the boundary condition
in the time direction from
Dirichlet to Neumann, and so takes a D$p$-brane to an E$p$-brane.
The E-branes preserve 16 of the 32 supersymmetries of the type II*
theories.
Smearing these solutions in the time direction gives the time-independent
solutions given by \ebr\ with $H$ given by \hre, and other solutions can be
obtained by smearing in spacelike directions.


\newsec{Brane Solutions of Arbitrary Signature}

For theories in spacetimes of signature $(S,T)$, we
are interested in generalised brane solutions with metric of the
form
\eqn\eki{ds^2=H^{-\aa } \eta _{ab} dX^a dX^b + H^\bb \ti \eta _{ij} dY^i dY^j,}
where $\eta _{ab}$ is a flat metric of signature $(s,t)$ and
$\ti \eta _{ij}$ is a flat metric of signature $(\tilde s, \tilde t
)=(S-s,T-t)$.
We will require $H$ to be a function of the transverse coordinates
$Y^i$ and find that the field equations imply that $H(Y)$   has to be a
harmonic function,
satisfying
\eqn\ekii{\ti \eta ^{ij} \pa _i \pa _j H=0,}
and determine the constants $\aa,\bb$.
The longitudinal space has signature $(s,t)$, and we refer to it as an
$(s,t)$-brane, so that a conventional
$p$-brane of a
Lorentzian theory with $(S,T)=(D-1,1)$ is a $(p,1)$-brane.
These solutions also have a non-vanishing $n$-form gauge field $C_n$ with
$n=s+t$
\eqn\ekiid{C_n= H^\gamma dX^1 \wedge \dots \wedge dX^n}
for some constant $\gamma$.

Different types of solutions arise for different choices of harmonic function.
A simple choice is the wave-type solution
\eqn\ekiii{H=A \sin (K_iY^i+c), \qq \ti \eta ^{ij}K_iK_j=0}
with $K$ a constant null vector.
For the  Euclidean transverse space (as in $p$-branes) there are no non-trivial
null vectors $K$, but there are non-trivial solutions
if the transverse space has both spacelike and timelike dimensions.
Here we shall concentrate on solutions of the form
\eqn\eha{ H=c+ {Q\over \tt ^{m}}}
or
\eqn\ehas{H=c+ {Q'\over \ss ^{m}}}
where $m=S+T-s-t-2$, $Q,Q'$ are real constants and
$\tt, \ss $ are the proper time and distance defined by
\eqn\abcde{\ss^2 =  -\tt ^2=  \ti \eta _{ij} Y^i Y^j.}
These solutions correspond  to a source located at a point in the transverse
space-time.
The two solutions \eha,\ehas\ are related by $Q'=i^m Q$ and so are distinct
real solutions
only
for odd $m$.
There are also multi-centre generalisations with
\eqn\multic{ H=c+ \sum _I {Q_I\over [\ti \eta _{ij}
(Y^i-Y^i_I)(Y^j- Y_I^j)]^{m/2}} }
corresponding to sources at the points $Y^i_I$ in the transverse space.

The single-centre solutions have   potential  singularities on the \lq
light-cone'
$\ss^2=0$,
where $H$ diverges, and on the hyperboloid where $H=0$.
As in the case of the E-branes, the
regions  $\ss^2\ge 0$  and $\ss^2 \le 0$ corresponding to the exterior and
interior
of the light-cone
define two distinct   solutions and in each case we take the sign of
$Q$ or $Q'$
so that
$H>0$ in the appropriate region. We will sometimes refer to the $(s,t)$-brane
solution
with metric \eki\ with \eha\ or \ehas\ and coordinates restricted to $\s^2 \ge
0$  as an $(s,t,+) $
brane and that for the region
$\s ^2\le 0$ as an $(s,t,-)$ brane.
 Then for the  $(s,t,-) $ solution with $\tt^2 \ge 0$, we take
\eha\ with $c\ge 0$, $Q>0$ while for the  $(s,t,+) $ solution with
$\ss^2\ge 0$ we take \ehas\
with $c\ge 0$, $Q'>0$; for $m$ even, this is of course equivalent to
taking
\eha\ with $c\ge 0$, $Q<0$.
We will be
particularly interested in the asymptotic form of the geometries near
$\ss^2=0$.
Note that this asymptotic form can be different for the two solutions
$(s,t,\pm) $ \us.
We will focus on the cases in which this is a product
of constant curvature spaces (generalising the usual $p$-brane case in which it
is a product of
a sphere and an anti-de Sitter space) in which the brane solution interpolates
between
this asymptotic geometry and flat space.

M-theory is the strong coupling limit of the type IIA string \wit\ and is a
theory in 10+1 dimensions
whose low-energy effective field theory is 11-dimensional
supergravity \cjs\ with bosonic action
\eqn\mmm{S_M=\int d^{11} x \sqrt{-g} \left( R - {G_4^2\over 48}
\right) -{1\over 12} \int C\wedge G \wedge G.}

The M2-brane solution of the 10+1 supergravity action \mmm\ is given by
\dufstel\
\eqn\mtwo{\eqalign{ds^2=&H^{-2/3} (-dt^2+dx_1^2+dx_2^2)
+ H^{1/3}(dy_3^2 + \dots + dy_9^2+dy_{10}^2),\cr
C_{012}&=H^{-1},\cr}}
where $H(y_3,\dots ,y_{10})$ is a harmonic function
in the transverse space. For a single membrane at $y=0$, we take
\eqn\fdkfo{H=1+{Q\over y^6},}
 where here and throughout $y^2=\sum_i y_i^2$, where $i$ runs over
the spatial indices in the transverse space.
The world-volume has signature (2,1).
The M2-brane solution has bosonic symmetry $ISO(2,1) \times SO(8)$.
It is nonsingular at $y=0$ with near-horizon geometry
$AdS_4 \times S^7$ \inter.

The   M5-brane   \guven\ is given by
\eqn\mfive{\eqalign{ds^2&=H^{-1/3} (-dt^2+dx_1^2 + \dots  +dx_5^2)
+H^{2/3} (dy_6^2 + dy_7^2+dy_8^2+dy_9^2+dy_{10}^2),\cr
&C_{t12345}=H^{-1},\cr}}
where
\ek{H=1+{Q \over y^3}}
 and $y^2=y_6^2+y_7^2 + \dots + y_{10}^2$.
This solution has bosonic symmetry $ISO(5,1) \times SO(5)$
and interpolates between $AdS_7 \times S^4$ and flat space \inter.

We will now consider the analogues of the M2-brane and  M5-brane
that occur in the M* theory.
The M* theory
\christwo\ is  the strong coupling limit of  the IIA* theory and is a theory
in 9+2 dimensions
whose field theory limit is a supergravity theory with bosonic action \chrisone
\eqn\mstar{S_M=\int d^{11} x \sqrt{g} \left( R + {G_4^2\over 48} \right)
-{1\over 12} \int C \wedge G \wedge G.}
Note that the sign of the kinetic term of  $G_4$ is opposite to that of the
action \mmm;
as was shown in \us, the sign of the kinetic term is intimately
related with the
world-volume signatures that can occur.
The M* theory with action  \mstar\ has  brane
solutions
with world-volume signatures (3,0) and (1,2) \us,   while  if
the sign of the
kinetic term of $G_4$ had been the opposite of that in \mstar\ to give a
Lagrangian
$R-  {G_4^2/ 48}+...$ in 9+2 dimensions, there would have been a membrane
solution
with 2+1
dimensional world-volume.
The sign of the $G_4$ kinetic term in actions \mmm\ and \mstar\ is
determined by supersymmetry \christwo.

The (1,2)-brane of the M*-theory is given by
\eqn\msonetwo{\eqalign{ds^2=&H^{-2/3} (-dt^2-dt'^2+dx^2)
+ H^{1/3}(dy_2^2 + \dots + dy_9^2),\cr
C_{tt'x}&=H^{-1},\cr}}
where $H(y_3,\dots ,y_{10})$ is again a  harmonic function in $\R^8$,
which we can take to be \fdkfo. The world-volume has signature
(1,2), with two times. This solution has bosonic symmetry
$ISO(1,2) \times SO(8)$. The transverse space is Euclidean, so there is only
one solution, the
(1,2,+) solution. Near $y=0$, the metric takes the form
\eqn\msonetwomet{ds^2 = {U^2\over R^2} (-dt^2-dt'^2+dx^2) + {R^2dU^2\over U^2}
+4 R^2d\Omega_7^2,}
which is the metric on $AAdS_4\times S^7$, where
$d\Omega_n^2$ is the metric on the $n$-sphere of unit volume,
$AAdS_4$ is the
de Sitter-like space of signature 2+2 given  by the coset $SO(3,2)/SO(2,2)$
\us,
  $U=Q^{-1/6}y^2/2$ and
$R=Q^{1/6}/2=R_{S^7}/2$.
The (1,2)-brane interpolates between the flat space $\R^{9,2}$ and
$AAdS_4\times S^7$.

The second membrane-type solution of the M* theory is the (3,0)-brane given by
\eqn\msthreezero{\eqalign{ds^2=&H_2^{-2/3} (dx_1^2+dx_2^2+dx_3^2)
+ H_2^{1/3}(-dt^2-dt'^2+dy_4^2 + \dots + dy_9^2),\cr
C_{123}&=H_2^{-1},\cr}}
where $H$ is a harmonic function  on the transverse space.
The world-volume is Euclidean, with
signature (3,0), and has isometry group $ISO(3)\times SO(6,2)$.
There are two distinct complete solutions, the $(3,0,+)$ brane given by
\msthreezero\ in the region
$\sigma^2=y^2-t^2-t'^2 \ge 0$ and the $(3,0,-)$ brane given  by
\msthreezero\ in the region
 $\sigma^2 \le 0$.
Near $y=0$ with
$\sigma^2 \ge 0$
the metric approaches   $H^4 \times AAdS_7$ as $y \to 0$, while for  $\sigma^2
\le 0$ it approaches
  $dS_4 \times AdS_7$. Hence the $(3,0,+)$-brane    solution    interpolates
between the flat space $\R^{9,2}$ and $H^4 \times AAdS_7$ while  the
$(3,0,-)$ brane    solution
 interpolates
between the flat space $\R^{9,2}$ and  $dS_4 \times AdS_7$ \us.

The M* theory has a  (5,1)-brane  solution (analogous to the M5-brane of
M-theory)
which is given by
\eqn\mfivestar{\eqalign{ds^2&=H^{-1/3} (-dt^2+dx_1^2 + \dots  +dx_5^2)
+H^{2/3} (-dt'^2 +dy_6^2 + dy_7^2+dy_8^2+dy_9^2),\cr
&C_{t12345}=H^{-1},\cr}}
Here $H$ is a harmonic function and
 there are two solutions.
In the $(5,1,-)$ solution
\eqn\ewtrw{H=1+{Q\over\tau^3},}
where $Q>0$ and the transverse coordinates are restricted to the region
 $\tau^2=y^2-t'^2 > 0$, where   $y^2=y_6^2+y_7^2 + \dots + y_9^2$.
In the (5,1,+) solution
\eqn\ufgjh{H=1+{Q\over\sigma^3},}
where the transverse coordinates are restricted to the region
 $\sigma^2=t'^2-y^2>0$, with $Q>0$.
Both solutions have bosonic symmetry
$ISO(5,1) \times SO(4,1)$ and world-volume signature (5,1).

There is a IIA string theory in a spacetime with signature 5+5 whose strong
coupling limit is the M$'$ theory with signature 6+5
\refs{\christwo}.
The field theory limit of M$'$ theory
is a supergravity theory in 6+5 dimensions with bosonic action
 \eqn\mprime{S_{M'}=\int d^{11} x \sqrt{-g} \left( R - {G_4^2\over 48}
\right) -{1\over 12} \int C\wedge G \wedge G.}
This has branes of world-volume signature 2+1, 0+3, 5+1, 3+3 and 1+5, and in
each case there are two
solutions $\ss^2 \ge 0$ or $\ss^2 \le 0$, except for the (1,5)-brane which has
a  transverse
space of Euclidean signature.

The various brane solutions in eleven dimensions are summarised in Table
1.

\vskip 0.5cm
{\vbox{
\begintable
  | $C_3$, $s+t=3$ | $\tilde C_6$, $s+t=6$ \elt
 $M_{10,1}$ | (2,1) | (5,1) \elt
 $M_{9,2}$ | $(3,0,\pm), (1,2) $| $(5,1,\pm) $\elt
 $M_{6,5}$ |$ (2,1,\pm), (0,3,\pm)$ | $(5,1,\pm), (3,3,\pm), (1,5)$
\endtable

{\bf Table 1} The M-branes with world-sheet signature $(s,t)$ coupling to
the 3-form gauge field $C_3$ or its 6-form dual $\ti C_6$ in the
various M-theories with signature
$(S,T)$. For Lorentzian transverse spaces, there are two solutions, $(s,t,\pm)$
}}
\vskip .5cm

The  brane solutions of the IIA-type and IIB-type
theories are summarised in Tables 2 and 3 respectively.
The type II branes all arise from solutions of the various 11-dimensional
theories \us.
The generalised fundamental strings have two-dimensional world-sheets of
signature $(s,2-s)$ and
couple to the NS-NS 2-form gauge field
$B_2$, while  the generalised NS 5-branes couple to its dual $\ti B_6$.
The branes coupling to the RR $n$-form gauge fields $C_n$
or their duals $\ti C_m$ are D-branes on which fundamental strings can end and
the solutions are of
the form \eki\ with $\aa =\bb = 1/2$.

We have included the D8-brane family of branes with $s+t=9$ that are obtained
by T-duality from the
results of \us. Note that the  $IIA_{5,5}^*$ theory, defined as the spacelike
T-dual
of the  $IIB_{5,5}^*$ theory or the timelike T-dual
of the  $IIB_{5,5}$ theory  \christwo, is related to the  $IIA_{5,5}$
theory by $g_{\mu \nu } \to -g_{\mu \nu }$,
interchanging space and time, and so every $(s,t)$ brane of the
$IIA_{5,5}$ theory
corresponds to a $(t,s)$   brane of the  $IIA_{5,5}^*$ theory. These
branes are needed in checking
  the T-duality relations of the various D-branes.
For example, a timelike T-duality takes the (3,3) brane of the $IIB_{5,5}$
theory
to a (3,4) brane or a (3,2) brane of the $IIA_{5,5}^*$ theory.

Note that each entry in tables 1,2,3  defines two solutions whenever the
transverse space has
indefinite signature, one with
$\ss^2\ge 0$ and one with $\ss^2\le 0$.

\vskip 1cm
{\vbox{
\begintable
  | $C_1$ | $B_2$ | $C_3$ | $\tilde C_5$ | $\tilde B_6$ | $\tilde
C_7$ | $\tilde
C_9$ \elt
 $IIA_{10,0}$ | (1,0) | (2,0) | -- | (5,0) | -- | -- | -- \elt
 $IIA_{9,1}$ | (0,1) | (1,1) | (2,1) | (4,1) | (5,1) | (6,1) | (8,1) \elt
 $IIA_{9, 1}^*$ | (1,0) | (1,1) | (3,0) | (5,0) | (5,1) | (7,0) | (9,0) \elt
 $IIA_{8,2}$ | (0,1) |
$\hbox{ (2,0),} \atop \hbox{(0,2)  }$
|$\hbox{ (3,0),} \atop \hbox{(1,2)  }$
 | (4,1) | (5,1) |
(7,0),(5,2) | (8,1)
\elt
 $IIA_{6,4}$ | (1,0) |$\hbox{ (2,0),} \atop \hbox{(0,2)  }$ |
$\hbox{ (2,1),} \atop \hbox{(0,3)  }$
 |
$\hbox{(5,0),(3,2),} \atop
\hbox{(1,4) }$| (5,1),(3,3) | (6,1),(4,3) | (5,4) \elt
 $IIA_{5,5}$ | (0,1) | (1,1) | $\hbox{ (2,1),} \atop \hbox{(0,3)  }$ |
$\hbox{(4,1),(2,3),} \atop
\hbox{(0,5)}$ | $\hbox{(5,1),(3,3),} \atop
\hbox{(1,5) }$| (4,3),(2,5) | (4,5)
\endtable

{\bf Table 2} The branes with world-sheet signature $(s,t)$ of the
various
$IIA_{S,T}$ theories with
signature
$(S,T)$, coupling to RR $n$-forms $C_n$, the NS-NS 2-form $B_2$ or its dual $
\tilde B_6$.
}}
 \vskip 1cm

{\vbox{
\begintable
  | $C_0$ | $B_2$ | $C_2$ | $ C_4$ | $\tilde B_6$ | $\tilde C_6$ |
$\tilde C_8$
\elt
 $IIB_{9,1}$ | -- |  (1,1) |  (1,1) | (3,1) | (5,1) | (5,1) | (7,1) \elt
 $IIB^*_{9,1}$ | (0,0) | (1,1) | (2,0) | (4,0) | (5,1) | (6,0) |
(8,0) \elt
 $IIB_{9, 1}'$ | (0,0) |  (2,0) |  (1,1) | (4,0) | (6,0) | (5,1) |
(8,0) \elt
 $IIB_{7,3}$ | -- |  $\hbox{(2,0),} \atop \hbox{(0,2) }$|
$\hbox{(2,0),} \atop
\hbox{(0,2)}$ |
(3,1),(1,3) | (6,0),(4,2) | (6,0),(4,2) |
$\hbox{(7,1),} \atop \hbox{(5,3) }$\elt
 $IIB_{5,5}$ | -- | (1,1) |  (1,1) |  (3,1),(1,3) |
$\hbox{(5,1),(3,3),} \atop
\hbox{(1,5)}$ |
$\hbox{(5,1),(3,3),} \atop \hbox{(1,5)}$ |
$\hbox{(5,3),} \atop \hbox{(3,5) }$\elt
 $IIB^*_{5,5}$ | (0,0) | (1,1) |  $\hbox{(2,0),} \atop \hbox{(0,2)}$ |
$\hbox{(4,0),(2,2),} \atop
\hbox{(0,4)}$ |
$\hbox{(5,1),(3,3),} \atop \hbox{(1,5)}$ | (4,2),(2,4) | (4,4) \elt
 $IIB_{5,5}'$ | (0,0) | $\hbox{ (2,0),} \atop \hbox{(0,2) }$|  (1,1) |
$\hbox{(4,0),(2,2),} \atop
\hbox{(0,4)}$ | (4,2),(2,4) | $\hbox{(5,1),(3,3),} \atop
\hbox{(1,5)}$ | (4,4)
\endtable

{\bf Table 3} The branes with world-sheet signature $(s,t)$ of the
various
$IIB_{S,T}$ theories with
signature
$(S,T)$.
}}

%
\newsec{Branes, Interpolations and De Sitter Spaces }

The M2-brane can be viewed as a soliton interpolating between 11-dimensional
Minkowski space and
the solution $AdS_4\times S^7$, while the M5-brane
interpolates  between 11-dimensional Minkowski space and
the solution $AdS_7\times S^4$ \inter.
There are two M-brane solutions corresponding  to  each entry in table 1
whenever
the transverse space has
indefinite signature,
one with
$\ss^2\ge 0$ and one with $\ss^2\le 0$, and the asymptotic form of the geometry
near $\ss^2=0$ was found to be
 a coset-space solution of signature $S+T$ for each case \us.
In each case, the metric is given by \eki\ with the harmonic function given
by \ehas\ with
vanishing constant piece, $c=0$.
The coset spaces that
arise are as follows:

\noindent {\bf I-Solutions of M-theory}

\noindent a-(2,1)-brane: $AdS_4 \times S^7 = {O(3,2)\over O(3,1)}
\times {O(8)\over O(7)}$.

\noindent b-(5,1)-brane: $AdS_7 \times S^4 = {O(6,2)\over O(6,1)} \times
{O(5)\over O(4)}$.

\noindent {\bf II-Solutions of M*-theory}

\noindent a-(1,2)-brane: $AAdS_4 \times S^7 = {O(3,2)\over O(2,2)} \times
{O(8)\over O(7)}$.

\noindent b-$(5,1,-)$-brane  and (3,0,+)-brane:
$AAdS_7 \times H^4 = {O(6,2)\over O(5,2)} \times {O(4,1)\over O(4)}$.

\noindent c-(5,1,+)-brane  and $(3,0,-)$-brane:
$AdS_7 \times dS_4 = {O(6,2)\over O(6,1)} \times {O(4,1)\over O(3,1)}$.

\noindent {\bf III-Solutions of M$'$-theory}

\noindent a-$(2,1,-)$-brane and (3,3,+)-brane:
$AAdS_4 \times {O(4,4)\over O(4,3)} = {O(3,2)\over O(2,2)}
\times {O(4,4)\over O(4,3)}$.

\noindent b-(2,1,+)-brane and, $(3,3,-)$-brane:
$AdS_4 \times {O(4,4)\over O(4,3)} = {O(3,2)\over O(3,1)}
\times {O(4,4)\over O(3,4)}$.

\noindent c-$(5,1,-)$-brane and (0,3,+)-brane:
$AAdS_7 \times -dS_4 = {O(6,2)\over O(5,2)} \times {O(1,4)\over O(1,3)}$.

\noindent d-(5,1,+)-brane, $(0,3,-)$-brane:
$AdS_7 \times -H^4 = {O(6,2)\over O(6,1)} \times {O(1,4)\over O(4)}$.

\noindent e-(1,5)-brane:
$-AAdS_7 \times S^4 = {O(2,6)\over O(2,5)} \times {O(5)\over O(4)}$.

For a given space $N$ of signature $(S,T)$, we denote by $-N$ the space of
signature
$(-S,-T)$ given by multiplying the metric by $-1$, so that whereas
$AdS_n$ is a space of signature $(n-1,1)$,
$-AdS_n$ is a space of signature $(1,n-1 )$.
In the cases IIb,c and IIIa-d, the same geometry arises as the asymptotic limit
for two different
brane solutions. We will discuss some of the implications of this
in section 9.

We will now extend this to the
brane solutions of the IIB theories with $s+t=4$
that correspond to the D3-brane that are listed in the $C_4$ column of table 3,
and find their
interpolating geometries.
The $(s,t)$ D-brane solutions are all of the form
\eqn\ekii{ds^2=H^{-1/2 } \eta _{ab} dX^a dX^b +
H^{1/2} \ti \eta _{ij} dY^i dY^j,}
where $\eta _{ab}$ is a flat metric of signature $(s,t)$ and
$\ti \eta _{ij}$ is a flat metric of signature $(S-s,T-t)$, with
  $H$   a function of the transverse coordinates
$Y^i$. For the D3-brane family for which $s+t=4$,
 the dilaton is zero and
the four-form antisymmetric tensor is
given
by $C_4 = -H^{-1}\epsilon $, where $\epsilon$ is the
volume-form on the longitudinal part of the spacetime.
 We define
$y^2 = \Sigma y_i^2$ and $\tilde t^2 = \Sigma \tilde t_j^2$ and let
$\sigma^2=y^2-\tilde t^2
= \ti \eta _{ij} Y^i Y^j $ and $\tau^2=\tilde t^2 - y^2=-\ti \eta _{ij} Y^i
Y^j $.
In this case, we take the harmonic function to be
\eqn\hfdslfh {H=1 + {Q^2\over \sigma^4 }}
so that $H\ge 1$ and
\eqn\sjffs{
H^{1/2} = \pm  {Q \over \sigma^2 }+...}
when $\ss^2$ is small.
When the transverse space has indefinite signature,
there are two distinct solutions, one defined
for  $\ss^2 \ge 0$ and  one for $\ss^2 \le 0$.
In each of the two
cases, $\ss^2 \ge 0$ and  $\ss^2 \le 0$, we choose the sign so that $H^{1/2}$
is positive in \sjffs.
The $(s,t)$-brane solution interpolates between
flat space and the asymptotic geometry as $\ss \to
0$, given by \ekii\ with $H=Q^2/\ss ^4$.
We list these asymptotic geometries below.

\vskip 0.5cm
{\vbox{

\begintable
String Theory |  $(s,t,\pm)$   | Asymptotic Geometry \elt
$IIB_{9,1}$ | (3,1)  |
$O(4,2)/O(4,1) \times O(6)/O(5)$ {}$=AdS_5\times S^5$
\elt
$IIB^*_{9,1}$, $IIB'_{9,1}$ | $(4,0,\pm)$|
$O(5,1)/O(5) \times O(5,1)/O(4,1)${}$=dS_5\times H^5$
 \elt
$IIB_{7,3}$ |$ (3,1,\pm) $| $O(4,2)/O(4,1) \times O(4,2)/O(3,2)$
{}$=AdS_5\times AAdS_5$
\elt
$IIB_{7,3}$ | (1,3,+)  | $O(2,4)/O(2,3)
\times O(6)/O(5)$ {}$={}-AAdS_5\times S^5$
\elt
$IIB_{5,5}$ | (3,1,+)  | $O(4,2)/O(4,1)
\times O(2,4)/O(1,4)$ {}$={}AdS_5\times - AdS_5$
\elt
$IIB_{5,5}$ | $(3,1,-) $ | $O(4,2)/O(3,2)
\times O(2,4)/O(2,3)$ {}$={}AAdS_5\times - AAdS_5$
\elt
$IIB_{5,5}$ | (1,3,+)  | $O(4,2)/O(3,2)
\times O(2,4)/O(2,3)$ {}$={}AAdS_5\times - AAdS_5$
\elt
 $IIB_{5,5}$ | $(1,3,-)$ | $O(4,2)/O(4,1)
\times O(2,4)/O(1,4)$ {}$={}AdS_5\times - AdS_5$
\elt
$IIB^*_{5,5}$, $IIB'_{5,5}$ | (4,0,+) |   $O(5,1)/O(5)
\times O(1,5)/O(5)${}$=H^5\times -H^5$
\elt
$IIB^*_{5,5}$, $IIB'_{5,5}$ | $(4,0,-) $|   $O(5,1)/O(4,1) \times
O(1,5)/O(1,4)${}$=dS^5\times -dS^5$
\elt
$IIB^*_{5,5}$, $IIB'_{5,5}$ | $(2,2,\pm) $  |  $O(3,3)/O(3,2) \times
O(3,3)/O(2,3)$
\elt
$IIB^*_{5,5}$, $IIB'_{5,5}$ | (0,4,+)   | $O(5,1)/O(4,1) \times
O(1,5)/O(1,4)${}$=dS^5\times -dS^5$
\elt
$IIB^*_{5,5}$, $IIB'_{5,5}$ |$ (0,4,-)$ |  $O(5,1)/O(5) \times
O(1,5)/O(5)${}$=H^5\times -H^5$
\endtable
{\bf Table 4} Asymptotic geometries near $\ss=0$ of the D-branes
with world-volume signature
$(s,t)$ with $s+t=4$ for the $(s,t,+)$ solutions with $\ss^2\ge 0$ and the
$(s,t,-)$ solutions with
$\ss^2\le 0$.}}
\vskip .5cm

\newsec{World-Volume Actions}

An $(s,t)$-brane is invariant under the Poincar\' e group
$ISO(s,t)$ and the
world-volume action should be a superPoincar\' e-invariant  theory  in $(s,t)$
dimensions with
16 supersymmmetries. In some cases the Poincar\' e symmetry will be
enhanced to
the conformal group $SO(s+1,t+1)$,
in which case the world-volume theory is invariant under the corresponding
superconformal
group with 32 supersymmetries. If the $(s,t)$-brane is embedded in $(S,T)$
dimensions, the
world-volume action
should also have an R-symmetry that contains $SO(\ti s , \ti t)$, where
\eqn\esti{\ti s =S-s, \qquad \ti t=T-t.}
Here and in the following sections, we will find the
field theories in $(s,t)$ dimensions with 16 supersymmetries that are the
unique candidates for
world-volume actions for the various branes, and postpone until section 9 a
discussion of the
differences between $(s,t,+)$ branes and $(s,t,-)$ branes.

The $(s,t)$-branes that couple to the Ramond-Ramond gauge fields in
a string
theory in $(S,T)$ dimensions with $S+T=10$
are generalised D-branes and the dynamics are governed by
fundamental strings ending on the brane. The low-energy
effective action
for these strings is a world-volume super-Yang-Mills (SYM) theory
(with a Born-Infeld-type action) in $(s,t)$ dimensions.
In 10 dimensions, the only signatures $(S,T)$ that allow
an $N=1$ superalgebra with 16 supersymmetries are those admitting
Majorana-Weyl
spinors, and these are the signatures (9,1) and (5,5) (together with the
reverse signature (1,9)).
The (9,1) SYM theory can be dimensionally reduced on a Euclidean
$p$-torus $T^p$ to
give the usual maximal SYM theory in $(9-p,1)$ dimensions with
R-symmetry $SO(p)$
or on a Lorentzian torus $T^{p,1}$ to give a Euclidean SYM theory
in $(9-p,0)$
dimensions with R-symmetry $SO(p,1)$.
Similarly, the (5,5) SYM theory can be reduced on a torus $T^{p,q}$ with
signature $(p,q)$ (with $p\le 5, q\le 5$) to give a SYM theory in signature
$(5-p,5-q)$ with $SO(p,q)$ R-symmetry.

One apparent problem is that there are many more types of
generalised D-branes
in tables 2 and 3 than there are types of SYM theories, and it is not
clear whether
all cases can be covered by the available SYM theories. For
example,   an $(s,t)$
brane of the IIB theory in (7,3) dimensions has $ ISO(s,t)\times
SO(7-s,3-t)$
symmetry, and a natural way that this could arise would be from
compactifying a
Yang-Mills theory in (7,3) dimensions on a torus $T^{7-s,3-t}$, but
there is no
supersymmetric Yang-Mills theory in (7,3) dimensions. This means that the
world-volume theory must either come from a Yang-Mills theory in (7,3)
dimensions without supersymmetry, or must arise in some other way.
We will address this issue in the next section, and show that in each case
the world-volume theory is in fact a SYM theory in $s+t$ dimensions with these
symmetries, and
explain how this comes about.

The M-theories in (10,1), (9,2) and (6,5) dimensions have
generalised membranes
with $s+t=3$ whose world-volumes are theories with 8 scalars and 8
fermions with
Poincar\' e
symmetry $ISO(s,t)$ which is enhanced to the conformal group $SO(s+1,t+1)$
and R-symmetry $SO(S-s,T-t)$ at a conformal fixed point.
There are also generalised 5-brane solutions with world-volume signatures
$(s,t)$=(5,1), (3,3), (1,5) and these are described by tensor
multiplets in $(s,t)$ dimensions with a 2-form gauge-field with
self-dual field
strength, 5 scalars and 4 fermions with an
$SO(S-s,T-t)$ R-symmetry and $SO(s+1,t+1)$ conformal symmetry.
The generalised membranes will be discussed in section 7 and the generalised
5-branes in section 8.

In each case, each of these branes can be linked to standard branes of
M-theory or string theory by chains of dualities, and this implies
that for $N$
coincident branes there is a $U(N)$ gauge symmetry.

\newsec{World-Volume Theories for Generalised D-Branes}

In this section, we will discuss the world-volume theories of the
D-branes of the various type II string theories. These are the
branes that
couple to RR gauge fields and on which the fundamental string
(which has either
Lorentzian or Euclidean world-sheet, depending on the case) can end.

The usual type IIA and IIB theories in (9,1) dimensions have
D$p$-branes with
world-volume signature $(p,1)$ (with $p$ even for IIA and odd for IIB)
and these
have effective world-volume theories which are
SYM theories in $(p,1)$ dimensions obtained by reducing (9,1)
dimensional SYM
on the Euclidean torus $T^{9-p}$.
Timelike T--duality takes the type IIA and IIB theories   to the
IIB* and IIA*
theories in (9,1) dimensions
respectively, and takes a D-brane with Lorentzian world-volume of
signature
$(p,1)$
to an E-brane with Euclidean signature $(p,0)$.
The effective world-volume theory is Euclidean SYM in $(p,0)$
dimensions obtained
by reducing (9,1) dimensional SYM
on the Lorentzian torus $T^{9-p,1}$.
T-dualities take  D-branes or E-branes  in (9,1) dimensions to
D-branes or E-branes  in (9,1) dimensions.

Similarly, the IIA and IIB, IIA* and  IIB*
theories in (5,5) dimensions have D-branes with world-volume
signature $(s,t)$
and these
have effective world-volume theories which are
SYM theories in $(s,t)$ dimensions obtained by reducing (5,5)
dimensional SYM
on the torus $T^{5-s,5-t}$. (The IIA* theory in  (5,5) dimensions
is obtained
from the IIA theory by the mirror transformation $g_{\mu \nu} \to -g_{\mu \nu}$
interchanging space
and   time, and so
is equivalent to the IIA theory \christwo.)

Thus in each of these cases, the world-volume theory for an $(s,t)$
D-brane in
$(S,T)$ dimensions (which is (9,1) or (5,5)) is $(s,t)$-dimensional SYM
obtained by
reducing from $(S,T)$ SYM on $T^{\ti s, \ti t}$ to give a theory with
$ISO(s,t)\times SO(\ti s, \ti t)$ symmetry.
For example, the (2,1)-brane in (9,1) dimensions is a (2,1)-dimensional
SYM theory obtained by reducing from (9,1)-dimensional SYM theory on
$T^7$ to give a theory with $ISO(2,1) \times SO(7)$ symmetry, which
is enhanced to an $O(3,2) \times O(8)$ symmetry at the conformal fixed point.

As mentioned in the previous section, this pattern cannot carry
over to the
string theories in (8,2), (7,3) or (6,4) dimensions.
In each case, the brane is expected to have a symmetry containing
$ISO(s,t)\times SO(\ti s, \ti t)$, and this would be the symmetry
obtained by reducing a Yang-Mills theory in $(S,T)$ dimensions on
$T^{\ti s, \ti t}$. But
there is no SYM theory for these values of $(S,T)$, so that if this
picture is correct, there can be no conventional supersymmetry.

However, there is another way of getting a SYM theory with
the right  symmetry. Consider, for example, the $(3,1)$ brane of the
IIB${}_{7+3}$ theory, which should have
$ISO(3,1)\times SO(4,2)$ symmetry, and which could be expected to
be governed
by a SYM theory. If there were a SYM theory in (7,3) dimensions,
this could
arise by reducing on $T^{4,2}$, but there is no such theory.
However, reducing
the SYM theory in (5,5) dimensions on $T^{2,4}$ does give a
SYM theory in (3,1)  dimensions with the required $ISO(3,1)\times
SO(4,2)$
symmetry. However, this would give  an SYM action  in (3,1)
dimensions in
which 2 scalars have a kinetic term of the \lq right' sign, and 4
with the \lq
wrong' sign, while the (3,1) brane should have 4 scalars with a
kinetic term of
the \lq right' sign, which are the zero-modes for
translations in the transverse spatial dimensions,
and 2 with the \lq wrong' sign, which are the zero-modes for
translations in the transverse temporal dimensions.
However, the right scalar action could be obtained by starting from (5,5)
dimensional SYM in which the {\it whole action} is multiplied by
$-1$, so that
the  gauge fields   have kinetic terms of the \lq
wrong' sign;
this is clearly consistent with supersymmety.

This works for all the D-branes of the type II theories in (8,2),
(7,3) or
(6,4) dimensions.
In each case, there is a candidate world-volume SYM theory in $(s,t)$
dimensions
with $ISO(s,t)\times SO(\ti s, \ti t)$ symmetry which is obtained
by reducing
an SYM theory in $(\ti S, \ti T)$ dimensions on
$T^{ \ti t, \ti s,}$, where
\eqn\eerw{\ti S=s+\ti t,\qquad \ti T=t+\ti s}
For this to work, it is essential that $(\ti S, \ti T)$ should be
(9,1), (5,5)
or (1,9), so that an SYM theory in $(\ti S, \ti T)$ dimensions exists,
and remarkably this is the case for each D-brane in these theories, as is
easily checked.
The IIB$'$ theories in
(9,1) or
(5,5) dimensions, which are the strong-coupling duals of the IIB* theories,
are  related to
these theories by T-dualities (the  (9,1) IIB$'$ theory is T-dual to the (10,0)
and (8,2)
IIA theories while the   (5,5) IIB$'$ theory is T-dual to the (6,4)
IIA theory),
 and so the same pattern should persist for
these   theories
also; indeed, $(\ti S, \ti T)$ is (9,1), (5,5) or (1,9) for each of the
D-branes of  the IIB$'$ theories, as is required.
In each of these cases, the SYM action in  $(\ti S, \ti T)$
dimensions has  a gauge field kinetic term of the
wrong sign, with all other signs following from supersymmetry.
As another example, consider the (1,3)-brane solution of IIB$_{7+3}$.
The SYM theory in this case arises from the reduction of a (1,9)
SYM theory with the `wrong' sign on $T^{0,6}$ to give a $1+3$-dimensional
SYM theory with $ISO(1,3)\times O(6)$ symmetry. For completeness,
we note that the (3,1), (2,0), (4,2) and (5,3) solutions of
IIB$_{7+3}$ arise from reductions of `wrong' sign (5,5) SYM,
the (0,2), (6,0) and (7,1) solutions of IIB$_{7+3}$ arise from reductions
of `wrong' sign (9,1) SYM, and the (1,3) and (0,2)
solutions of IIB$_{7+3}$ arise from reductions
of `wrong' sign (1,9) SYM.

Thus we have found the following picture.
For the theories in which the fundamental string has a Lorentzian
world-sheet,
i.e. the IIA, IIB, IIA*, IIB* theories in (9,1) or (5,5) dimensions, the
world-volume theory is $(S,T)$ dimensional  SYM compactified on
$T^{\ti s, \ti t}$ to give a theory with $ISO(s,t)\times SO(\ti s, \ti t)$
symmetry, and the
$(S,T)$ dimensional SYM theory has action
\eqn\edftdsa{S=-\4 \int d^{10}x F^2+...}
with the \lq right' sign.

For the theories in which the fundamental brane has 2 Euclidean
dimensions,
i.e. the IIB$'$ theories in (9,1) or (5,5) dimensions, the IIA theories in
(10,0),(8,2) and (6,4) dimensions, and the
IIB theory in (7,3) dimensions, an SYM theory with the right symmetry is
obtained by
reducing the SYM theory in $(\ti S, \ti T)$ dimensions on
$T^{ \ti t, \ti s,}$,
where the $(\ti S, \ti T)$-dimensional SYM action has the \lq wrong' sign
for the gauge field kinetic term.
\eqn\edftda{S= \4 \int d^{10}x F^2+...}
For the D-branes in (10,0), (8,2), (7,3) or (6,4) dimensions, this
is the {\it unique } SYM theory with $ISO(s,t)\times SO(\ti s, \ti t)$
symmetry, and so
supersymmetry implies this must be the
correct world-volume theory in these cases, and then T-duality
implies that
this construction must also give the world-volume theory for the
D-branes of
the IIB$'$ theories.

As a further example, a (4,0) brane in (9,1) dimensions should be
governed by
an SYM theory in (4,0) dimensions with
$ISO(4)\times SO(5,1)$ symmetry, and such an SYM theory can be
obtained either
by reducing (9,1) dimensional SYM on $T^{5,1}$
or by reducing (5,5) dimensional SYM on $T^{1,4}$; the former gives the
world-volume theory of the (4,0) brane of the IIB*$_{9,1}$ theory,
and the
latter gives the world-volume theory of the (4,0) brane of the
IIB$'_{9,1}$
theory.

The D-brane world-volume theories are then always supersymmetric
Yang-Mills
theories with 16 supersymmetries, and the identifications of these
theories
above is consistent with T-duality.
The theories in which the fundamental string is Lorentzian have SYM
theories
with lagrangians $-F^2+...$ while those
in which the fundamental \lq string' is Euclidean have SYM theories with
lagrangians $+F^2+...$.

\subsec{Superconformal Symmetry}

For those D-branes with $s+t=4$ (including the usual D3-brane) the
SYM theory
is invariant under the conformal group $SO(s+1,t+1)$, containing the
Poincar\' e
group $ISO(s,t)$.
The bosonic symmetry is then
$SO(s+1,t+1)\times SO(\ti s, \ti t)$ and, as the theory is
supersymmetric,
there  must be
a superconformal group for which this is
 the bosonic part.
For the usual D3-brane, this is the group $SU(2,2/2)$, for the
E4-brane of the
IIB*$_{9+1}$ theory, this is $SU(2,2/2)^*$
and for the other cases other real forms of the same group emerge.
We summarise
the results below (recall that $SU(4)\sim SO(6),SU(2,2)\sim SO(4,2),
SL(4,\R)\sim SO(3,3)$).

\vskip 0.5cm
{\vbox{

\begintable
 Theory | $(s,t)$ | Bosonic Symmetry | Supergroup | SYM on $T^{p,q}$ \elt
 $IIB_{9,1}$ | (3,1) | $O(4,2)\times O(6)$ | SU(2,2/4) | $SYM_{9,1}$ on $T^{6}$
\elt
 $IIB^*_{9,1}$ | (4,0) | $O(5,1)\times O(5,1)$ | $SU^*(2,2/4)$ | $SYM_{9,1}$ on
$T^{5,1}$ \elt
 $IIB'_{9,1}$ | (4,0) | $O(5,1)\times O(5,1)$ | $SU^*(2,2/4)$ | $SYM_{5,5}$ on
$T^{1,5}$ \elt
 $IIB_{7,3}$ | (3,1) | $O(4,2)\times O(4,2)$ | SU(2,2/2,2)| $SYM_{5,5}$ on
$T^{2,4}$ \elt
 $IIB_{7,3}$ | (1,3) | $O(4,2)\times O(6)$ | SU(2,2/4) | $SYM_{1,9}$ on $T^{6}$
\elt
 $IIB_{5,5}$ | (3,1) | $O(4,2)\times O(4,2)$ | SU(2,2/2,2)| $SYM_{5,5}$ on
$T^{2,4}$ \elt
 $IIB_{5,5}$ | (1,3) | $O(4,2)\times O(4,2)$ | SU(2,2/2,2)| $SYM_{5,5}$ on
$T^{4,2}$ \elt
 $IIB^*_{5,5}$ | (4,0) | $O(5,1)\times O(5,1)$ | $SU^*(2,2/4)$ | $SYM_{5,5}$ on
$T^{1,5}$ \elt
 $IIB^*_{5,5}$ | (2,2) | $O(3,3)\times O(3,3)$ |  SL(4/4) | $SYM_{5,5}$ on
$T^{3,3}$ \elt
 $IIB^*_{5,5}$ | (0,4) | $O(5,1)\times O(5,1)$ | $SU^*(2,2/4)$ |  $SYM_{5,5}$
on $T^{5,1}$ \elt
 $IIB'_{5,5}$ | (4,0) | $O(5,1)\times O(5,1)$ | $SU^*(2,2/4)$ | $SYM_{5,5}$ on
$T^{1,5}$ \elt
 $IIB'_{5,5}$ | (2,2) | $O(3,3)\times O(3,3)$ | SL(4/4) | $SYM_{5,5}$ on
$T^{3,3}$ \elt
 $IIB'_{5,5}$ | (0,4) | $O(5,1)\times O(5,1)$ | $SU^*(2,2/4)$ | $SYM_{5,5}$ on
$T^{5,1}$
\endtable
{\bf Table 5} The symmetries of the D-branes  with world-volume
signature
$(s,t)$ with $s+t=4$ coupling to $C_4$ in the various IIB theories. The
world-volume theories arise
in each case from super-Yang-Mills theory in 9+1, 5+5 or 1+9 dimensions
compactified on a torus
$T^{p,q}$.}}
\vskip .5cm

The table also gives the massless sector of the world-volume theories.
In each case, it is
given by   the super-Yang-Mills theory in 9+1, 5+5 or 1+9 dimensions
dimensionally reduced on a torus
$T^{p,q}$. The sign of the kinetic term of the gauge field is the
\lq wrong' one ($+F^2$) for the
branes of the $IIB'$ theories and the $IIB_{7+3}$ theories.
For example, the (4,0)-brane of the $IIB^*_{5,5}$ has a world-volume
theory given by the SYM theory
in 5+5 dimensions   compactified on
$T^{1,5}$ to obtain a theory with manifest
$ISO(4)\times O(1,5)$   symmetry which is enhanced to
$O(5,1) \times O(1,5)$ by the conformal
invariance.

\newsec{The M2-Brane Family}

In this section we consider the world-volume theories of the
M-branes with
$s+t=3$ in $S+T=11$ dimensions  for  $S=10, 9$ or 6.
A vertical dimensional reduction in a space dimension then gives
the $(s,t)$
D-brane of a IIA theory in $(S-1,T)$
dimensions while a time reduction gives
the $(s,t)$ D-brane of a IIA theory in $(S-1,T)$ dimensions.
In either case, we know from the previous section which SYM theory with a
vector and 7 scalars gives the world-volume theory of the $(s,t)$
D-brane, and
dualising the vector to an eighth scalar then gives the
world-volume theory of
the M-brane, generalising the relation between the D2-brane and
the M2-brane \eleven.
The sign of the kinetic term of the extra scalar then follows from the
analysis of section 8 of \us.

It has been argued \seibergir\ that the world-volume theory of the
M2-brane has an infra-red fixed
point at which the theory becomes superconformally invariant.  We shall assume
that
all the world-volume theories in the M2-brane family have a superconformal
fixed point. This
assumption is supported by the fact that the near-horizon limit of each of
these branes is
superconformally invariant.

The results are as follows.
The world-volume theory is a theory in $(s,t)$ dimensions $(s+t=3)$
with 8 scalars and 8 fermions and a bosonic symmetry
$ISO(s,t)\times SO(\ti s, \ti t)$. The R-symmetry is $SO(\ti s, \ti
t)$ and
there are $\ti s$ scalars with the \lq right' sign kinetic term,
and $\ti t$
with the \lq wrong' one.
The theory is assumed to have a conformal fixed point at which the
$ISO(s,t)$ symmetry is enhanced to the conformal group
$SO(s+1,t+1)$, and the theory is invariant under a superconformal
group with
bosonic subgroup
$SO(s+1,t+1)\times SO(\ti s, \ti t)$.
For the usual M2-brane, this is the supergroup
$OSp(4/8)$ while for the other cases it is a different real form of
this group.

For example, the SYM
theory of the (2,1)-brane solution of (10,1) M theory
arises from the (9,1) SYM theory by compactifying on $T^7$. The $O(7)$
symmetry is enhanced to $O(8)$ by dualising the vector $A_\mu$ to an
eighth scalar $X^8$. The $ISO(2,1)\times O(8)$ symmetry is then
enhanced to $O(3,2) \times O(8)$ at the conformal fixed point.
The SYM
theory of the (3,0)-brane solution of the (9,2) M* theory
arises from the (9,1) SYM theory by compactifying on $T^{6,1}$. The $O(6,1)$
R-symmetry is enlarged to $O(6,2)$ by dualising the vector $A_\mu$ to an
eighth scalar $X^8$,
which
 has a kinetic term of the `wrong' sign. The
$ISO(3)\times O(6,2)$ symmetry is then enhanced to $O(4,1) \times O(6,2)$ at
the conformal point.

A similar analysis holds for the (2,1) solution of M*
and for the (2,1) and (0,3) solutions of the
(6,5) M$'$.
We summarise the  results below.

\vskip 0.5cm
{\vbox{

\begintable
 $(S,T)$ | $(s,t)$ | Bosonic Symmetry | Supergroup \elt
 (10,1) | (2,1) | $O(3,2)\times O(8)$ | OSp(4/8) \elt
 (9,2) | (3,0) | $O(4,1)\times O(6,2)$ | $OSp^*(4/8)$ \elt
 (9,2) | (1,2) | $O(3,2)\times O(8)$ | OSp(4/8) \elt
 (6,5) | (2,1) | $O(3,2)\times O(4,4)$ | OSp(4/4,4) \elt
 (6,5) | (0,3) | $O(4,1)\times O(6,2)$ | $OSp^*(4/8)$
\endtable
{\bf Table 5} The symmetries of the M-branes  with world-volume
signature
$(s,t)$
with $s+t=3$ coupling to
$C_3$  in the
various M-theories with signature
$(S,T)$.
}}
\vskip .5cm

\newsec{The M5-Brane Family}

In this section we consider the world-volume theories of the
M-branes with
$s+t=6$ in $S+T=11$ dimensions  for  $S=10, 9$ or 6.
A spatial double dimensional reduction of an $(s,t)$ M-brane gives an
$(s-1,t)$ D-brane in the IIA theory in signature  $(S-1,T)$
and a timelike double dimensional reduction gives an
$(s,t-1)$ D-brane in the IIA theory in signature $(S,T-1)$. In
either case, the
world-volume theory of the D-brane is the SYM theory in
$(s-1,t)$ dimensions or $(s,t-1)$ dimensions
determined in section 5, with bosonic symmetry
$ISO(s-1,t) \times SO(\ti s, \ti t)$  or $ISO(s,t-1) \times SO(\ti
s, \ti t)$.
The M-brane world-volume theory is a self-dual tensor multiplet
with 5 scalars
and 4 spinor fields
that gives the SYM theories on spacelike or timelike dimensional
reduction.
This tensor multiplet theory has bosonic symmetry
containing $ISO(s,t) \times SO(\ti s, \ti t)$, with the $ \ti s+\ti t=5$
scalars in a vector representation of $SO(\ti s, \ti t)$, so that
there are $\ti s$ scalars with a kinetic term of the right sign
and $\ti t $
with the wrong sign.
This theory is in fact conformally invariant, so that the
Poincar\' e symmetry $ISO(s,t)$ is enhanced to the
 conformal group $SO(s+1,t+1)$, and the theory is invariant under a
superconformal group  with bosonic subgroup
$SO(s+1,t+1)\times SO(\ti s, \ti t)$, and in each case is some
real form of $OSp (8/4)$.

For example,
  the M5-brane or
(5,1)-brane of the (10,1) M theory
 arises
as the strong coupling limit of the D4-brane.
The D4-brane world-volume theory is (4,1)-dimensional SYM with $ISO(4,1)\times
O(5)$.
At strong coupling an extra dimension emerges, so that the Poincar\' e symmetry
becomes
$ISO(5,1)$ and is a subgroup of the conformal symmetry $SO(6,2)$,
so that the full
bosonic symmetry is $O(6,2) \times O(5)$.
On the other hand, the (5,1)-brane of the (9,2) M* theory
is the strong-coupling limit of the E5-brane of the IIA* theory in 9+1
dimensions, and the
E5-brane symmetry $O(4,1)\times ISO(5)$  is enhanced to
$O(4,1) \times ISO(5,1)$ in the  strong coupling
limit in which an extra time dimension emerges, and conformal invariance then
increases this to the $O(4,1) \times O(6,2)$ symmetry of a self-dual tensor
theory in 5+1
dimensions with $O(4,1)$ R-symmetry and twisted supersymmetry with 16
supercharges. A similar
analysis holds for the other M5-branes. We summarise the  results below:

\vskip 0.5cm
{\vbox{

\begintable
 $(S,T)$ | $(s,t)$ | Bosonic Symmetry | Supergroup \elt
 (10,1) | (5,1) | $O(6,2)\times O(5)$ | $OSp(4/6,2)$ \elt
 (9,2) | (5,1) | $O(6,2)\times O(4,1)$ | $OSp^*(4/8)$ \elt
 (6,5) | (5,1) | $O(6,2)\times O(4,1)$ | $OSp^*(4/8)$ \elt
 (6,5) | (3,3) | $O(4,4)\times O(3,2)$ | OSp(4/4,4) \elt
 (6,5) | (1,5) | $O(6,2)\times O(5)$ | $OSp^*(4/8)$
\endtable
{\bf Table 6} The symmetries of the M-branes  with world-volume
signature
$(s,t)$
with $s+t=5$ coupling to
$\ti C_6$  in the
various M-theories with signature
$(S,T)$.
}}
\vskip .5cm

\newsec{ Dualities Between Conformal Field Theories and De Sitter Space
Theories.}

The massless sector of the D3-brane, M2-brane and M5-brane world-volume
theories
are  field theories in 3+1, 2+1 and 5+1 dimensions which are
superconformally invariant and
are dual to the IIB string theory in $AdS_5\times S^5$,
M-theory in
$AdS_4\times S^7$ and
M-theory in
$AdS_7\times S^4$, respectively \mal.
The original argument leading to this holographic duality was based on
considering a certain limit
of the brane
theory.
In \mal,   $N$ parallel D$3$-branes separated by distances of order $\rr$
were considered and
the zero-slope limit
$\aa '\to 0$  was taken keeping $r=\rr / \aa '$ fixed, so that the energy of
stretched strings
remained finite. This decoupled the bulk and string degrees of freedom  leaving
a theory on the
brane which is $U(N)$ ${\cal N}=4$ \sym\ with Higgs expectation values, which
are of order $r$,
corresponding to the brane separations.   The corresponding D$3$-brane
supergravity solution is of
the form \dbr\ and
has charge $Q=a^2 \propto Ng_s/{\aa '}^2$ where $g_s$ is the string coupling
constant, which is
related to the \sym\ coupling constant $g_{YM}$ by $g_s=g_{YM}^2$.
Then as $\aa'\to 0$, $Q$ becomes
large and the background becomes $AdS_5\times S^5$. The IIB string theory in
the $AdS_5\times S^5$
background is a good description if the curvature $R \sim 1/a^2$ is not too
large, while if $a^2$ is
large, the \sym\ description is reliable.  In the 't Hooft limit in which $N$
becomes large while
$g_{YM}^2N$ is kept fixed, $g_s \sim 1/N$, so that as $N\to \infty$, we get the
free string limit
$g_s\to 0$, while string loop corrections correspond to $1/N$ corrections in
the \sym\ theory.
Similar arguments apply to the M2 and M5 brane cases.

This was extended to the case of E4-branes in \chrisone, and the two
E$4$-brane solutions, the $(4,0,\pm)$-branes,  correspond  to whether the
separation
between the
E$4$-branes that is kept fixed is spacelike or timelike. Recall that the
scalars of the \sym\ theory
are in a vector representation of the $SO(5,1)$ R-symmetry, where those in the
{\bf 5} of
$SO(5)\subset SO(5,1)$ have kinetic terms of the right sign and correspond  to
brane separations in
the 5 spacelike
 transverse dimensions, while the remaining ($U(N)$-valued) scalar  is a ghost
and corresponds to
timelike separations of the E-branes.
For the case of   $N$ parallel E$4$-branes of the IIB* string theory
separated by
distances of order $\rr$  in one of the 5 spacelike transverse dimensions, we
take the zero-slope
limit
$\aa '\to 0$ keeping $\ss=\rr / \aa '$ fixed.  This gives a decoupled theory on
the brane consisting of the $U(N)$
${\cal N}=4$ Euclidean
\sym, with Higgs expectation values of order $\ss$ for the scalars
corresponding to the spacelike
separations.  The corresponding supergravity background is the $(4,0,+)$-brane.
 We again
have  $Q=a^2 \propto Ng_s/{\aa '}^2$ and $g_s=g_{YM}^2$, so that for
large $N$, the system can be
described by the IIB* string theory in $dS_5\times H^5$ if $a^2$ is large
and by the large $N$
Euclidean \sym\ theory when
$a^2$ is small. In the 't Hooft limit,  string loop corrections again
correspond to $1/N$
corrections in the \sym\ theory.
For $N$ E$4$-branes of the IIB* string theory separated by distances of
order $T$
in the timelike transverse dimension,
we take the zero-slope limit
$\aa '\to 0$ keeping $\tt=T/ \aa '$ fixed.
This gives a decoupled theory on the brane consisting of the $U(N)$ ${\cal
N}=4$ Euclidean \sym,
 with Higgs expectation values of order $\tt$ for the scalars   corresponding
to the timelike
separations.  The corresponding supergravity background is the $(4,0,-)$ brane.
 Again
 for large $N$, the
system can be  described by the IIB* string theory in $dS_5\times H^5$ if
$a^2$ is large and by
the large $N$ Euclidean \sym\ theory when
$a^2$ is small.

Similar arguments
can be given for each of the brane solutions in the D3-brane, M2-brane and
M5-brane families with
various   signatures.
Considering $N$ parallel $(s,t)$ branes
and taking the Maldacena limit, we formally obtain a duality
between the $(s,t)$ brane world-volume superconformal field theory
 and the IIB or M-theory in $(S,T)$ dimensions in the
spacetime given by the
asymptotic form of the $(s,t)$ brane
solution.
Moreover, it is straightforward to check that the superconformal group found
for the world-volume
theories matches the super-isometry group of the asymptotic geometry, so that
the symmetries of the
two theories agree.

However, there are some new complications that arise for the cases considered
here.
The metric \eki,\eha\ corresponds to {\it two} solutions, the $(s,t,+)$ brane
and the $(s,t,-)$ brane
 (unless the transverse space is Euclidean), and each of these has a different
asymptotic
geometry in general. Although for a given $(s,t)$ there are    two  brane
solutions, there is only {\it
one}  $(s,t)$ worldvolume field theory allowed by supersymmetry.
A given spacetime can arise as the   asymptotic geometry for several different
types of brane.
For example,  the  M*-theory  solution  $AdS_7\times dS_4$ is the asymptotic
geometry for both
the (5,1,+) brane and the $(3,0,-)$ brane of M*-theory.
The asymptotic geometry is often the product of two non-compact spaces, and so
holography might be expected to operate
in each factor, with a field theory associated with some surface
(e.g. the boundary) for each.
Clearly, the dualities between theories will be more complicated than in the
cases considered in \mal.

Let us consider the example of the (0,3) brane of M$'$ theory in some detail.
The transverse space has signature (6,2) and the world-volume theory has 8
scalars transforming as
a vector under the O(6,2) R-symmetry and 8 fermions transforming as a spinor of
O(6,2).
The ISO(3) Poincar\' e symmetry is enhanced to O(1,4) conformal symmetry,
and the superconformal group is $OSp^*(4/8)$.
 In the usual way,
for $N$ branes the world-volume fields take values in the Lie algebra of $U(N)$
and giving
some of the scalars
expectation values corresponds to separating the branes. Two branes at two
positions in $\R ^{6,2}$
have a separation which can be spacelike, timelike or null.

The two  associated supergravity solutions, the (0,3,+) brane and the
$(0,3,-)$ brane, both have isometry
$ISO(3)\times O(6,2)$. The asymptotic geometry for the   (0,3,+) brane is
$AAdS_7\times -dS_4$
while that for the $(0,3,-)$ brane is $AdS_7\times -H_4$, and both have
isometry group
$O(4,1)\times O(6,2)$, contained in the super-isometry group $OSp^*(4/8)$.
If the two solutions, the $(0,3,\pm)$ branes, are both associated with the same
field theory, then
this would suggest that M$'$ theory in  $AdS_7\times -H_4$ is dual to
M$'$ theory in $AAdS_7\times -dS_4$.
Moreover, precisely the same asymptotic geometries arise for the
(5,1) brane -- the (5,1,+) brane tends to  $AdS_7\times -H_4$ and the
$(5,1,-)$ to
$AAdS_7\times -dS_4$. This would then lead to the suggestion that
the $(5,1,\pm)$   theory should be dual to  the
$(0,3,\mp)$ theory.
Then there are four theories -- the world-volume theories of the (5,1) and
(0,3) branes, and M$'$
theory in the backgrounds $AdS_7\times -H_4$ and  $AAdS_7\times -dS_4$ -- each
with
the same symmetry group, $OSp^*(4/8)$, and naive application of standard
arguments suggest they
might all be related by dualities.

To see what is going on, it will be useful to consider the application of the
Maldacena argument to this case.
Consider a two-centre (0,3) brane solution
with metric \eki\ and harmonic function
\eqn\multica{ H=c+   {Q \over (\ti \eta _{ij} Y^iY^j)  ^{3}}
+  {q \over [\ti \eta _{ij} (Y^i-Y^i_0)(Y^j- Y_0^j)]  ^{3}}}
corresponding to a brane of charge $Q$ at $Y=0$ and one of charge $q$ at
$Y=Y_0$.
Suppose further that $Q >> q$, corresponding to $N$ branes at
$Y=0$ and $n$ branes at $Y=Y_0$
with $N >> n$.
The contribution of the $n$ branes will be small except in a
neighbourhood of $Y_0$, and outside this neigbourhood the solution is
approximately that of $N$ branes at the origin, so that the $n$ branes can be
thought of as probes in the $N$ brane background.
If $Y_0$ is spacelike, we should take this $N$ brane geometry to be the (0,3,+)
solution while if it
is timelike, we should take the $(0,3,-)$ one, and more generally for a
multi-centre
(0,3,+) solution the positions $Y_0$ in \multic\ should all be spacelike, while
for the $(0,3,-)$
solution they should be timelike.
 These correspond to the spacelike and timelike interpolations of \chrisone.

This suggests that the (0,3) world-volume theory has (at least) two \lq
branches', one in which the
scalar expectation values are all spacelike, and one in which they are all
timelike.
Then the (0,3,+) brane solutions should correspond to the branch
of the world-volume theory in which the expectation values of the scalars are
all spacelike
and the  $(0,3,-)$ brane solutions to the   timelike branch.
However, these two branches are clearly part of the same connected moduli space
of one theory, so
that presumably one can move continuously between them.

We now take the  Maldacena limit in which the Planck length $l_p$ tends to zero
while the
brane
separation $Y_0$ scales so that $Y_0/l_p^3$ stays fixed.
For the case with $Y_0$ spacelike, this gives the  (0,3,+)
  asymptotic geometry, $AAdS_7\times -dS_4$, while for the timelike case
this gives $AdS_7\times -H_4$.
Then M$'$ theory in these two asymptotic geometries should each, by the usual
arguments, be
holographically related to the corresponding branch of the (0,3) world-volume
theory, and so M$'$
theory in these two different backgrounds correspond to different points in
the same connected
moduli space, and so in this sense are dual to each other.

Consider the space $AdS_7\times -H_4$. The hyperbolic space $-H_4$ has a
boundary
$-S^3$ and the (0,3) brane world-volume theory is associated with this
boundary, while
$ AdS_7$ has a timelike
boundary $S^5\times S^1$ (or $S^5\times\R$ for the covering space $CAdS_7$,
conformal
  to Minkowski space $\R^{5,1}$) and this boundary is associated with the
(5,1) brane world-volume theory.

For supersymmetric solutions of any of these theories of the form
$AdS_7\times Y_4$
with $Y^4$ compact (as in the solution of M-theory with $Y_4=S^4$),
there is a holographic relation between the theory in $AdS_7\times Y_4$ and a
5+1 dimensional CFT
on the  boundary of the anti-de Sitter space. The relation is that of
\refs{\witt,\holog};
if the fields bulk $\phi^i$ tend to prescribed functions $\phi_0(x)$ on the
$AdS$ boundary, the
partition function $Z(\phi_0^i)$ for the bulk theory  is identified with (a
generating functional for)
correlation functions  of a conformal field theory on the boundary.
Similarly, for solutions $X_7 \times \pm H_4$ with $X_7$ compact, then the
bulk partition
function with prescribed boundary values on the boundary $S^3$ of $H^4$ is
identified with
correlation functions of a Euclidean CFT on $S^3$ (since $H^4$ is the analytic
continuation of
$AdS_4$). Now the same should apply  if $X$ or $Y$ is non-compact, provided
trivial boundary
conditions are imposed on $X$ or $Y$.
In particular, for $AdS_7\times -H_4$, the M$'$ theory partition function
with prescribed boundary
conditions on $AdS_7$ and trivial boundary conditions on $-H_4$ should be
identified with
correlation functions of the $(5,1,+)$ brane
world-volume CFT, while
the M$'$ theory partition function with prescribed boundary
conditions on  $-H_4$ and trivial boundary conditions on $AdS_7$
 should be identified with
correlation functions of the $(0,3,-)$ brane
world-volume CFT.
When there are non-trivial boundary conditions for both spaces, it seems
unlikely
that the system could be described by either the 3-dimensional  CFT or the
6-dimensional CFT
on their own, but that perhaps both would be needed.

This suggests the following picture for the general case.
For a given $(s,t)$, the $(s,t,\pm)$ branes will tend to the asymptotic
geometries
\eqn\asdfg{
(s,t,+)-brane \to X^+_{(s+1,t)}\times
Y^+_{(\ti s -1,  \ti t)},
\qquad
(s,t,-)-brane \to X^-_{(s ,t+1)}\times
Y^-_{(\ti s,  \ti t  -1)} }
for   spaces $X,Y$, listed in section 4,
where $ X^+_{(s+1,t)}$ has signature $(s+1,t)$ etc.
In some situations, as in M-theory or the IIB string, one of the two cases will
be
absent.
The $(s,t)$ brane world-volume CFT will correspond to the bulk theory on
$X^\pm \times Y^\pm$
with prescribed boundary conditions on   $X^\pm$ and trivial  boundary
conditions on   $Y^\pm$.
The world-volume theory of the $(s,t)$ brane has two branches,
the  $(s,t,+)$ branch with   scalar expectation values in a spacelike direction
in the moduli space, and the $(s,t,-)$ branch with   scalar expectation values
in a timelike direction.
The $(s,t,+)$ branch of the field theory is
\lq dual' to the theory on $ X^+_{(s+1,t)}\times
Y^+_{(\ti s -1,  \ti t)}$ and is
associated with the boundary of $X^+_{(s+1,t)}$, while the
$(s,t,-)$ branch of the field theory is
\lq dual' to the theory on $X^-_{(s ,t+1)}\times
Y^-_{(\ti s,  \ti t  -1)} $
 and is
associated with the boundary of $X^-_{(s ,t+1)}$.
Further, assuming that  it is valid to view the
  $(s,t,\pm)$ branches as two regions in the modulli space of a single
theory, namely the $(s,t)$-brane world-volume theory, then this theory is
holographically related to the string or M-type theory on the two spaces
$X^+_{(s+1,t)}\times
Y^+_{(\ti s -1,  \ti t)}$ and $X^-_{(s ,t+1)}\times
Y^-_{(\ti s,  \ti t  -1)} $.
These two solutions also have a holographic dual associated with the boundary
of $Y$,
a conformal field theory in  $(\ti s -1,  \ti t-1)$ dimensions    associated
with the boundaries of both $
Y^+_{(\ti s -1,  \ti t)}$ and
of
$Y^-_{(\ti s,  \ti t  -1)}$, with the two branches of the field theory
corresponding to the two different solutions.
Then there is a quartet of related theories, the string or M-type theory on the
two spaces $X^+_{(s+1,t)}\times
Y^+_{(\ti s -1,  \ti t)}$ and $X^-_{(s ,t+1)}\times
Y^-_{(\ti s,  \ti t  -1) }$, and the conformal field theories  in  $(s,t)$
dimensions  and in $(\ti s -1,  \ti t-1)$ dimensions.
The relation between these theories that is suggested  is intriguing and
clearly deserves further investigation.

\vskip .5cm

\vskip1truecm

\noindent
{\bf Acknowledgements:}

CMH is supported by an EPSRC Senior Fellowship and would like to
thank Rockefeller University and Baruch College for hospitality.
RRK is supported by NSF Grant 9900773 and by a Eugene Lang Junior Faculty
Research Scholarship.

\listrefs
\end